\documentclass[12pt]{article}

\usepackage{fancybox}

\usepackage{cite}
\usepackage{float}
\usepackage{amsfonts}
\usepackage{amsmath}
\usepackage{amsbsy}
\usepackage{graphicx}
\usepackage{amssymb}
\usepackage{amsthm}
\usepackage{bm}
\usepackage{epsfig}
\usepackage{latexsym}
\usepackage{pdflscape}
\usepackage{color}
\numberwithin{equation}{section}

\allowdisplaybreaks

\DeclareMathOperator{\tr}{tr}

\newcounter{aff}

\begin{document}

\begin{titlepage}
\begin{flushright}
{\footnotesize YITP-14-98}
\end{flushright}
\begin{center}
{\Large\bf Exact Instanton Expansion of Superconformal\\[6pt]
 Chern-Simons Theories from Topological Strings}

\bigskip\bigskip
{\large 
Sanefumi Moriyama\footnote[1]{\tt moriyama@math.nagoya-u.ac.jp}
\quad and \quad
Tomoki Nosaka\footnote[2]{\tt nosaka@yukawa.kyoto-u.ac.jp}
}\\
\bigskip
${}^{*}$\,
{\small\it Kobayashi Maskawa Institute
\& Graduate School of Mathematics, Nagoya University\\
Nagoya 464-8602, Japan}
\medskip\\
${}^{*\dagger}$\,
{\small\it Yukawa Institute for Theoretical Physics,
Kyoto University\\
Kyoto 606-8502, Japan}
\end{center}

\begin{abstract}
It was known that the ABJM matrix model is dual to the topological string theory on a Calabi-Yau manifold.
Using this relation it was possible to write down the exact instanton expansion of the partition function of the ABJM matrix model.
The expression consists of a universal function constructed from the free energy of the refined topological string theory with an overall topological invariant characterizing the Calabi-Yau manifold.
In this paper we explore two other superconformal Chern-Simons theories of the circular quiver type.
We find that the partition function of one theory enjoys the same expression from the refined topological string theory as the ABJM matrix model with different topological invariants while that of the other is more general.
We also observe an unexpected relation between these two theories.
\end{abstract}

\begin{figure}[h]
\centering
\includegraphics[width=7cm]{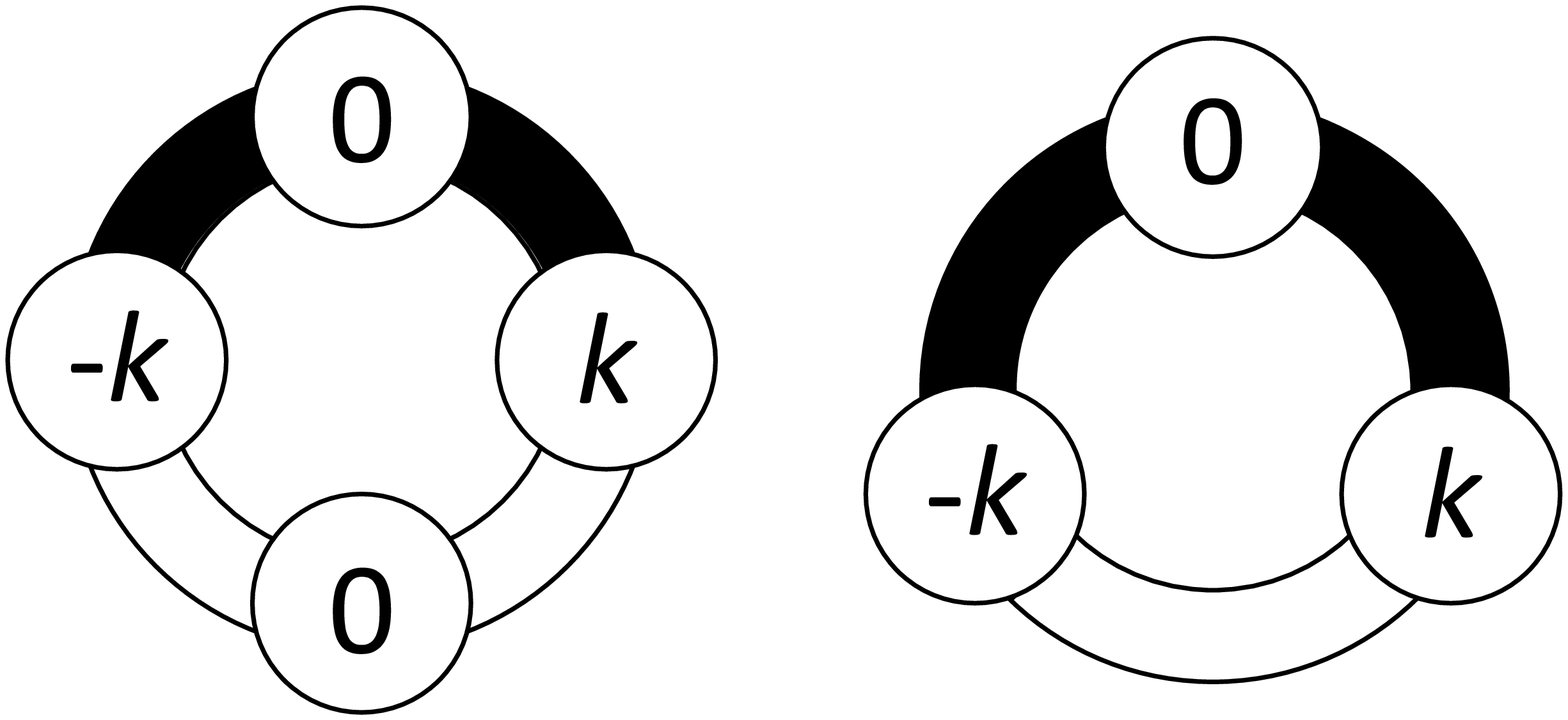}
\end{figure}

\end{titlepage}
\tableofcontents

\section{Introduction}\label{introduction}
Chern-Simons theory plays a central role in modern string theory.
It was known more than two decades ago that the Chern-Simons theory can be regarded as the topological string theory \cite{W}.
Interestingly, the relation to the topological string theory also appears in a supersymmetric case.
The ${\cal N}=6$ superconformal Chern-Simons theory with gauge group $U(N)_k\times U(N)_{-k}$ and bifundamental matters was proposed as the worldvolume theory of $N$ multiple M2-branes on the geometry ${\mathbb C}^4/{\mathbb Z}_k$ \cite{ABJM}.
With the help of the localization theorem which reduces the infinite dimensional path integral to a finite dimensional matrix integration, the partition function of this theory on $S^3$ is reduced to a matrix model \cite{KWY}, which we will call here the ABJM matrix model.
The ABJM matrix model was found to be dual to the topological string theory on a Calabi-Yau manifold, local $\mathbb{P}^1\times \mathbb{P}^1$ \cite{MPtop}.

After a series of studies \cite{DMP1,DMP2,FHM,MP,KEK,HMO1,PY,HMO2,CM,HMO3,HMMO}, finally the exact instanton expansion of the ABJM matrix model was written down \cite{HMMO}.
It is worthwhile to note that, in each step of the progress, the relation to the topological string theory played an essential role.
In \cite{DMP1} the leading large $N$ behavior $N^{3/2}$ \cite{KT} of the free energy was found from the relation.
Then, the all genus partition function was summed up to the Airy function \cite{FHM} by using the holomorphic anomaly equation \cite{BCOV} of the topological string theory on local ${\mathbb P}^1\times{\mathbb P}^1$ \cite{DMP1,DMP2}.
After taking care of the constant map \cite{KEK} and moving to the dual grand potential \cite{MP},
\begin{align}
J(\mu)=\log\Biggl[\sum_{N=0}^\infty Z(N)e^{\mu N}\Biggr],
\label{Jdef}
\end{align}
with the chemical potential $\mu$, again the numerical results of the worldsheet instanton part ($\sim e^{-\mu/k}$) \cite{HMO2} was compared with the free energy of the topological string theory on local ${\mathbb P}^1\times{\mathbb P}^1$.
Finally, the membrane instanton part ($\sim e^{-\mu}$) was once again determined by the Nekrasov-Shatashvili limit \cite{NS} of the free energy of the refined topological string theory \cite{HMMO}.

Aside from the perturbative part which is dual to the Airy function, the non-perturbative part of the grand potential was found to be \cite{HMMO} ($s_L=2j_L+1$, $s_R=2j_R+1$)\footnote{
Note that the grand potential $J(\mu)$ hereafter is slightly different from its original definition \eqref{Jdef} (denoted by $J^\text{periodic}(\mu)$ in this footnote) which is periodic in $\mu$, $J^\text{periodic}(\mu+2\pi i)=J^\text{periodic}(\mu)$.
The relation is given by
\begin{align}
e^{J^\text{periodic}(\mu)}=\sum_{n=-\infty}^\infty e^{J(\mu+2\pi in)}.
\end{align}
See \cite{HMO2} for more details.
}
\begin{align}
&J^\text{np}(\mu)
=\sum_{j_L,j_R}\sum_d\sum_{\sum_i{\bf d}_i=d}
N^{{\bf d}}_{j_L,j_R}
\nonumber\\&\quad\times
\sum_{n=1}^\infty\Biggl[
\frac{s_R\sin 2\pi g_sns_L}
{n(2\sin\pi g_sn)^2\sin 2\pi g_sn}
e^{-ndT_\text{eff}}
+\frac{\partial}{\partial g_s}
\biggl(g_s
\frac{-\sin\frac{\pi n}{g_s}s_L\sin\frac{\pi n}{g_s}s_R}
{4\pi n^2(\sin\frac{\pi n}{g_s})^3}
e^{-nd\frac{T_\text{eff}}{g_s}}\biggr)\Biggr].
\label{Jtop}
\end{align}
Here $N^{\bf d}_{j_L,j_R}$ is the BPS index \cite{HK,CKK} on local ${\mathbb P}^1\times{\mathbb P}^1$ with degree ${\bf d}$ and spin $(j_L,j_R)$, though for simplicity we only consider the diagonal case with all of the K\"ahler parameters $T_\text{eff}$ identical.
We identify the string coupling constant as $g_s=2/k$ and the effective K\"{a}hler parameter as $T_\text{eff}=4\mu_\text{eff}/k\pm\pi i$, where the relation between the effective chemical potential $\mu_\text{eff}$ and the original one $\mu$ was known explicitly for integral $k$ \cite{HMO3}.

So far we have explained how the relation between the ABJM matrix model and the topological string theory helps us in solving the ABJM matrix model.
Taking the relation reversely, we can regard the matrix model as the non-perturbative definition of the topological string theory.
It was noted \cite{HMMO} that in \eqref{Jtop} the topological information of the background geometry, local ${\mathbb P}^1\times{\mathbb P}^1$, is encoded solely in the BPS index $N^{{\bf d}}_{j_L,j_R}$ and that all the poles appearing in the universal function multiplied by $N^{{\bf d}}_{j_L,j_R}$ cancel among themselves provided $s_L+s_R+1\equiv 0\mod 2$ \cite{HMO2,CM,HMMO}.
Therefore, in \cite{HMMO} the expression \eqref{Jtop} was proposed as the non-perturbative completion of the topological string theory on an {\it arbitrary} background geometry.
Namely, if we want to consider the topological string theory on other backgrounds, all we have to do is to replace $N^{{\bf d}}_{j_L,j_R}$ by the BPS index on that background. 

From this viewpoint, however, it is unclear whether we can regard the expression \eqref{Jtop} as the non-perturbative completion of the topological string theory on an arbitrary background geometry, if the ABJM matrix model is the only example which has the topological string interpretation.
In other words, it is natural to ask whether, and how, the variation of the background is realized on the Chern-Simons theory side.
To answer this question, we shall explore other superconformal Chern-Simons theories.

\begin{figure}
\centering
\includegraphics[width=7cm]{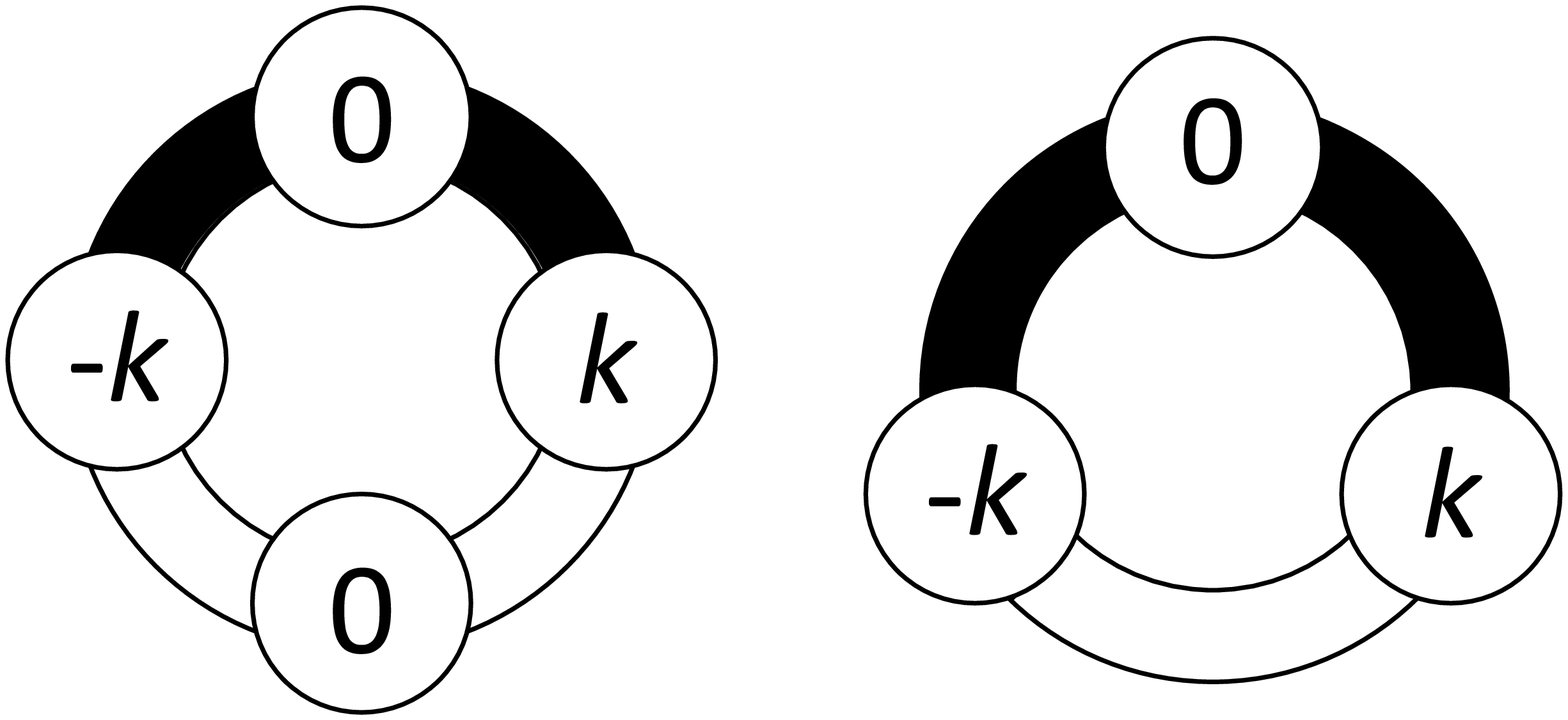}
\caption{
The circular quiver of the $(2,2)_k$ model characterized by $\{s_a\}_{a=1}^4=\{+1,+1,-1,-1\}$ (left) and that of the $(2,1)_k$ model characterized by $\{s_a\}_{a=1}^3=\{+1,+1,-1\}$ (right).
The sign $s_a$ is associated to the edge between the $a$-th vertex and $(a+1)$-th vertex (numbered counterclockwise).
The black and white colors are assigned to the edges with $s_a=+1$ and those with $s_a=-1$, respectively.
}
\label{22and21quiver}
\end{figure}

In an attempt of generalizations, let us consider the ${\cal N}=3$ superconformal Chern-Simons theories \cite{GY,JT} with gauge group $\prod_{a=1}^MU(N)_{k_a}$ ($\sum_ak_a=0$) and bifundamental matters between $U(N)_{k_a}$ and $U(N)_{k_{a+1}}$, which were built on the previous works \cite{KL,KLL}.
For this class of ${\cal N}=3$ superconformal theories, the grand potential defined in \eqref{Jdef} can be expressed as that of an ideal Fermi gas system \cite{MP}
\begin{align}
J(\mu)=\tr\log(1+e^{\mu-\widehat H}),
\label{JFG}
\end{align}
with the one-particle Hamiltonian $\widehat H$.
Furthermore, it was found in \cite{IK4} that if the levels are given by
\begin{align}
k_a=\frac{k}{2}(s_a-s_{a-1}),\quad s_a=\pm 1,
\end{align}
the supersymmetry is enhanced to ${\cal N}=4$.
In \cite{MN1}, we proposed to start with the study of the special cases where $s_a=+1$ and $s_a=-1$ are well separated
\begin{align}
\{s_a\}=\{\underbrace{+1,+1,\cdots,+1}_{q},
\underbrace{-1,-1,\cdots,-1}_{p}\},
\label{sofminimal}
\end{align}
and called the corresponding matrix models the $(q,p)_k$ models.
As examples, in figure \ref{22and21quiver} we display the quivers of the $(2,2)_k$ model and the $(2,1)_k$ model.
For the $(q,p)_k$ model, the one-particle Hamiltonian is especially simple,
\begin{align}
e^{-\widehat H}=\biggl[2\cosh\frac{\widehat Q}{2}\biggr]^{-q}
\biggl[2\cosh\frac{\widehat P}{2}\biggr]^{-p},
\end{align}
where the coordinate operator $\widehat Q$ and the momentum operator $\widehat P$ satisfy the canonical commutation relation $[\widehat Q,\widehat P]=i\hbar$ with $\hbar=2\pi k$.
Since it was known \cite{MP} that the perturbative part of the grand potential of this theory is
\begin{align}
J^\text{pert}(\mu)=\frac{C}{3}\mu^3+B\mu+A,
\label{Jpert}
\end{align}
and the explicit form of $C$ was also known \cite{HKPT,GHP,MP}, in \cite{MN1} we further proceeded to compute $B$ for general ${\cal N}=4$ theories, conjecture $A$ for the $(q,p)_k$ models and see the first few instantons for the $(2,1)_k$ model.

To seek the theories in which the instanton effects of the grand potential has the similar structure as \eqref{Jtop}, the $(2,2)_k$ model  in figure \ref{22and21quiver} would be the best one to start with among ${\cal N}=3$ theories for the following two reasons.
First, for this theory we already know the perturbative part of the grand potential explicitly, which we have to subtract first to investigate the non-perturbative effects.
Second, since the membrane instanton in the $(q,p)_k$ model consists of three sectors of $e^{-\frac{2\mu}{q}}$, $e^{-\frac{2\mu}{p}}$ and $e^{-\mu}$ as found in \cite{MN2}, it is expected that some special simplification occurs at $(q,p)=(2,2)$ where all the three exponents coincide.

Although it is straightforward to generalize our analysis of the non-perturbative effects in the ABJM matrix model to the $(q,1)_k$ model \cite{GM,HaOk,MN1}, it is not so trivial whether the study of the $(2,2)_k$ model is possible.
In the analysis of the $(q,1)_k$ model, it was important that the matrix element of the density matrix defined by
\begin{align}
\rho(Q_1,Q_2)\simeq\langle Q_1|e^{-\widehat H}|Q_2\rangle,
\label{density}
\end{align}
up to a similarity transformation introduced to make it hermitian, takes the form
\begin{align}
\rho(Q_1,Q_2)=\frac{E(Q_1)E(Q_2)}{M(Q_1)+M(Q_2)},
\label{rho}
\end{align}
for some functions $M(Q)$ and $E(Q)$.
Due to a lemma\footnote{
Interestingly, a similar structure is found in the Neumann matrices of the light-cone string field theory.
See e.g.\ (C.3) in \cite{GSB}.} in \cite{TW}, this structure allows us to compute $\tr\rho^n$ without difficulty, as we shortly review in section \ref{q1}.
Then, as in \cite{HMO1,PY,HMO2}, we can compute the exact values of the partition function $Z(N)$ up to a certain large number of $N_\text{max}$ and read off the coefficients of the grand potential $J(\mu)$.
In the $(q,2)_k$ model, however, the density matrix does not take the form of \eqref{rho}.
So it is unclear whether we can repeat the same analysis.

In this paper we shall answer these questions positively.
Namely, we show that in principle we can generalize our analysis to all of the $(q,p)_k$ models by a slight modification of \eqref{rho}.
After that, we concentrate on the $(2,2)_k$ model and show that the instanton expansion has exactly the same structure as \eqref{Jtop}.
We identify the BPS indices as in table \ref{BPS}.
In terms of the diagonal Gopakumar-Vafa invariants $n^d_g$, which do not distinguish one of the spins in the BPS indices,
\begin{align}
\sum_{\sum_i{\bf d}_i=d}\sum_{j_L,j_R}N^{\bf d}_{j_L,j_R}
\frac{s_R\sin 2\pi g_ss_L}{\sin 2\pi g_s}
=\sum_{g=0}^\infty n_g^d(2\sin\pi g_s)^{2g},
\label{NtoGVinv}
\end{align}
the results are listed in table \ref{GV}.
It is interesting to observe that the diagonal Gopakumar-Vafa invariants listed in table \ref{GV} match with those of the local $D_5$ del Pezzo geometry (see table 6 in \cite{KKV}), though the BPS indices $N^{\bf d}_{j_L,j_R}$ look different (see section 5.4 in \cite{HKP}).
\begin{table}
\begin{center}
\begin{tabular}{|c||c|}
\hline
$\;d\;$
&$\pm\sum_{\sum_i{\bf d}_i=d}\sum_{j_L,j_R}
N^{\bf d}_{j_L,j_R}(j_L,j_R)$\\
\hline\hline
$1$&$8(0,\frac{1}{2})$\\
\hline
$2$&$8(0,\frac{1}{2})+(0,\frac{3}{2})$\\
\hline
$3$&$8(0,\frac{1}{2})+8(0,\frac{3}{2})$\\
\hline
$4$&$(4+2m_1+5m_2)(0,\frac{1}{2})
+(30-m_1-m_2)(0,\frac{3}{2})+(9-m_2)(0,\frac{5}{2})$\\
&$+(5-3m_1-5m_2)(\frac{1}{2},0)
+m_1(\frac{1}{2},1)+m_2(\frac{1}{2},2)$\\
\hline
$5$
&$(-80+2m_3+5m_4+7m_5)(0,\frac{1}{2})
+(80-m_3-m_4)(0,\frac{3}{2})$\\
&$+(80-m_4-m_5)(0,\frac{5}{2})+(16-m_5)(0,\frac{7}{2})$\\
&$+(96-3m_3-5m_4-7m_5)(\frac{1}{2},0)
+m_3(\frac{1}{2},1)+m_4(\frac{1}{2},2)+m_5(\frac{1}{2},3)$\\
\hline
\end{tabular}
\caption{The BPS indices identified for the $(2,2)_k$ model.
$m_1,m_2,\cdots,m_5$ are some numbers which we cannot fix in our analysis.}
\label{BPS}
\end{center}
\end{table}
\begin{table}
\begin{center}
\begin{tabular}{|c||c|c|c|c|c|c|c|}
\hline
$d$&1&2&3&4&5&6&7\\
\hline\hline
$n_0^d$&$16$&$-20$&$48$&$-192$&$960$&$-5436$&$33712$\\
\hline
$n_1^d$&$0$&$0$&$0$&$5$&$-96$&$1280$&$-14816$\\
\hline
$n_2^d$&$0$&$0$&$0$&$0$&$0$&$-80$&$2512$\\
\hline
$n_3^d$&$0$&$0$&$0$&$0$&$0$&$0$&$-160$\\
\hline
$n_4^d$&$0$&$0$&$0$&$0$&$0$&$0$&$0$\\
\hline
\end{tabular}
\caption{The diagonal Gopakumar-Vafa invariants identified for the $(2,2)_k$ model.}
\label{GV}
\end{center}
\end{table}

After studying the $(2,2)_k$ model we revisit the $(2,1)_k$ model whose studies were initiated in \cite{MN1}.
Unexpectedly, we find that the worldsheet instanton part of the grand potential of the $(2,1)_k$ model is related to that of the $(2,2)_k$ model.

The organization of this paper is as follows.
In the next section we shall explain how the techniques used to study the $(q,1)_k$ models actually work for general $(q,p)_k$ models.
Using these techniques combined with the results from the WKB expansion we proceed to study the $(2,2)_k$ model and the $(2,1)_k$ model in section \ref{22&21}.
Finally in section \ref{discuss} we conclude with some future directions.

\bigskip

\noindent
{\large\bf Note added}

After this paper was submitted to arXiv, the authors of \cite{HHO} share with us their results of the WKB expansion for the $(2,1)_k$ model up to ${\cal O}(k^{29})$ along the line of \cite{CM,MN2}.
The series expansions are all consistent with our proposed function forms \eqref{a21}, \eqref{21der} and \eqref{cb} in section \ref{21mb}.

\section{Exact computation of partition functions}
In the previous works \cite{HaOk,MN1}, it was found that the density matrix has the special structure\footnote{Though it is not relevant to our current analysis, this structure of the resolvent is related to the integration equations in the thermodynamic Bethe ansatz.} not only for the ABJM $(1,1)_k$ model but also for the $(q,1)_k$ models.
It provides an efficient way to calculate the quantity $\tr\rho^n$, by which we can immediately obtain the exact values of the partition function $Z(N)$ as
\begin{align}
Z(1)=\tr\rho,\quad
Z(2)=-\frac{1}{2}\tr\rho^2+\frac{1}{2}(\tr\rho)^2,\quad
Z(3)=\frac{1}{3}\tr\rho^3-\frac{1}{2}(\tr\rho)(\tr\rho^2)
+\frac{1}{6}(\tr\rho)^3,
\end{align}
and so forth, according to the expressions of the grand potential \eqref{Jdef} and \eqref{JFG}.

In this section, after reviewing the techniques used for the $(q,1)_k$ model, we shall explain how a similar structure appears in the $(q,2)_k$ model, so that we can continue our analysis in a parallel manner.
We also shortly note that a similar analysis works for the general $(q,p)_k$ model as well.

\subsection{$(q,1)_k$ model}\label{q1}
Before going on to the $(q,2)_k$ model, we shall first review the structure of the density matrix $\rho$ in the $(q,1)_k$ model and the calculation of $\tr\rho^n$ with it.

Let us start with the density matrix \eqref{density} for the $(q,p)_k$ model
\begin{align}
\rho(Q_1,Q_2)
=\frac{1}{2\pi}\frac{1}{\bigl(2\cosh\frac{Q_1}{2}\bigr)^{q/2}}
\langle Q_1|\frac{1}{\bigl(2\cosh\frac{\widehat P}{2}\bigr)^p}|Q_2\rangle
\frac{1}{\bigl(2\cosh\frac{Q_2}{2}\bigr)^{q/2}},
\label{rhoq1q2}
\end{align}
with the matrix element given by
\begin{align}
\langle Q_1|\frac{1}{\bigl(2\cosh\frac{\widehat P}{2}\bigr)^p}|Q_2\rangle
=\int\frac{dP}{2\pi k}\frac{e^{i(Q_1-Q_2)P/\hbar}}
{\bigl(2\cosh\frac{P}{2}\bigr)^p}.
\label{rhoFourier}
\end{align}
For the $(q,1)_k$ model, using the Fourier transformation formula
\begin{align}
\int\frac{dP}{2\pi}\frac{e^{i(Q_1-Q_2)P/\hbar}}{2\cosh\frac{P}{2}}
&=\frac{1}{2\cosh\frac{Q_1-Q_2}{2k}},
\end{align}
we end up with
\begin{align}
\rho(Q_1,Q_2)
=\frac{1}{2\pi k}
\frac{1}{\bigl(2\cosh\frac{Q_1}{2}\bigr)^{q/2}}
\frac{1}{2\cosh\frac{Q_1-Q_2}{2k}}
\frac{1}{\bigl(2\cosh\frac{Q_2}{2}\bigr)^{q/2}}.
\end{align}
This takes the form of \eqref{rho} with the identifications
\begin{align}
M(Q)=2\pi ke^{\frac{Q}{k}},\quad
E(Q)=\frac{e^{\frac{Q}{2k}}}{\bigl(2\cosh\frac{Q}{2}\bigr)^{q/2}}.
\end{align}

Using the structure \eqref{rho}, we can calculate the powers of the density matrix $\rho^n$ as follows.
First, let us rewrite \eqref{rho} schematically as
\begin{align}
\{M,\rho\}=E\otimes E,
\label{rhocomm}
\end{align}
by regarding $\rho$, $M$ and $E$ respectively as a symmetric matrix, a diagonal matrix and a vector whose components are given by $(\rho)_{Q,Q^\prime}=\rho(Q,Q^\prime)$, $(M)_{Q,Q^\prime}=M(Q)\delta(Q-Q^\prime)$ and
$(E)_Q=E(Q)$ and performing the matrix product by an integration with respect to $Q$.
Using \eqref{rhocomm} repetitively, we arrive at the expression
\begin{align}
\{M,\rho^n]=\sum_{m=0}^{n-1}
(-1)^m(\rho^m\cdot E)\otimes (\rho^{n-1-m}\cdot E).
\end{align}
Here on the left-hand side we employ the anti-commutator for odd $n$ and the commutator for even $n$.
On the right-hand side both the multiplication among $\rho$ and that between $\rho$ and $E$ are performed by the integration though we insert a dot only for the latter one.
If we define
\begin{align}
\phi_m(Q)=\frac{(\rho^m\cdot E)(Q)}{E(Q)},
\label{phim}
\end{align}
the power $\rho^n$ is given by
\begin{align}
\rho^n(Q_1,Q_2)=\frac{E(Q_1)E(Q_2)}{M(Q_1)-(-1)^nM(Q_2)}
\sum_{m=0}^{n-1}
\phi_m(Q_1)\phi_{n-1-m}(Q_2).
\label{trrhon}
\end{align}

Here comes the important point of this formula.
Typically when we compute the power $\rho^n$ we have to multiply matrices $n$ times iteratively.
The formula \eqref{trrhon} states that, however, $\rho^n$ can be computed by picking up a specific vector $E$ and multiplying $\rho$ to it recursively as \eqref{phim}.
Hence, the formula \eqref{trrhon} substantially simplifies the computation.

\subsection{$(q,2)_k$ model}\label{q2tba}
Now we shall see how this trick works for the $(q,2)_k$ model.
Again using the Fourier transformation formula
\begin{align}
\int\frac{dP}{2\pi}\frac{e^{i(Q_1-Q_2)P/\hbar}}
{\bigl(2\cosh\frac{P}{2}\bigr)^2}
&=\frac{1}{2\pi k}\frac{Q_1-Q_2}{2\sinh\frac{Q_1-Q_2}{2k}}
\end{align}
in \eqref{rhoFourier}, we find that the density matrix \eqref{rhoq1q2} becomes
\begin{align}
\rho(Q_1,Q_2)=\frac{1}{(2\pi k)^2}
\frac{1}{\bigl(2\cosh\frac{Q_1}{2}\bigr)^{q/2}}
\frac{Q_1-Q_2}{2\sinh\frac{Q_1-Q_2}{2k}}
\frac{1}{\bigl(2\cosh\frac{Q_2}{2}\bigr)^{q/2}}.
\end{align}
If we introduce
\begin{align}
M(Q)=(2\pi k)^2e^{\frac{Q}{k}},\quad
E(Q)=\frac{e^{\frac{Q}{2k}}}{\bigl(2\cosh\frac{Q}{2}\bigr)^{q/2}},
\end{align}
this density matrix is written as
\begin{align}
\rho(Q_1,Q_2)=\frac{(Q_1-Q_2)E(Q_1)E(Q_2)}{M(Q_1)-M(Q_2)}.
\end{align}

Schematically, this result can be rewritten as
\begin{align}
[M,\rho]=(EQ)\otimes E-E\otimes(EQ).
\end{align}
Note that the multiplication $EQ$ is simply the multiplication as functions and should be regarded as a vector independent of $E$.
This means that the only difference from the $(q,1)$ model is that in this case we need to introduce two vectors correspondingly,
\begin{align}
\phi_m(Q)=\frac{(\rho^m\cdot E)(Q)}{E(Q)},\quad
\psi_m(Q)=\frac{(\rho^m\cdot EQ)(Q)}{E(Q)},
\end{align}
with which $\rho^n$ is written as
\begin{align}
\rho^n(Q_1,Q_2)=\frac{E(Q_1)E(Q_2)}{M(Q_1)-M(Q_2)}
\sum_{m=0}^{n-1}\bigl[\psi_m(Q_1)\phi_{n-1-m}(Q_2)
-\phi_m(Q_1)\psi_{n-1-m}(Q_2)\bigr].
\end{align}

To summarize, the computations needed to obtain the partition function $Z(N)$ are the following integrations:
the integrations which give the two series of vectors $\phi_m$, $\psi_m$ recursively as
\begin{align}
\phi_m(Q)&=\int dQ'\frac{1}{E(Q)}\rho(Q,Q')E(Q')\phi_{m-1}(Q'),
\quad
\phi_0(Q)=1,
\nonumber\\
\psi_m(Q)&=\int dQ'\frac{1}{E(Q)}\rho(Q,Q')E(Q')\psi_{m-1}(Q'),
\quad
\psi_0(Q)=Q,
\end{align}
and the trace
\begin{align}
\tr\rho^n=\int dQ\frac{E(Q)^2}{{dM}/{dQ}}
\sum_{m=0}^{n-1}\Bigl[\frac{d\psi_m(Q)}{dQ}\phi_{n-1-m}(Q)
-\frac{d\phi_m(Q)}{dQ}\psi_{n-1-m}(Q)\Bigr].
\end{align}

\subsection{$(q,p)_k$ model}
Before closing this section, we briefly explain how the above technique works for general $(q,p)_k$ models.
The Fourier transformation of $\bigl(2\cosh\frac{P}{2}\bigr)^{-p}$ for general $p$ is given as
\begin{align}
\int\frac{dP}{2\pi k}
\frac{e^{i(Q_1-Q_2)P/\hbar}}{\bigl(2\cosh\frac{P}{2}\bigr)^p}=
\begin{cases}
\displaystyle\frac{1}{2(p-1)!\cosh\frac{Q_1-Q_2}{2k}}
\prod_{j=1}^{\frac{p-1}{2}}
\biggl[
\biggl(\frac{Q_1-Q_2}{2\pi k}\biggr)^2+\frac{(2j-1)^2}{4}
\biggr]
&\;\text{for odd }p,\\
\displaystyle\frac{Q_1-Q_2}{4\pi k(p-1)!\sinh\frac{Q_1-Q_2}{2k}}
\prod_{j=1}^{\frac{p}{2}-1}
\biggl[
\biggl(\frac{Q_1-Q_2}{2\pi k}\biggr)^2+j^2
\biggr]
&\;\text{for even }p.
\end{cases}
\end{align}
From these formula it follows that, with $M(Q)\propto e^{\frac{Q}{k}}$, $\{M,\rho\}$ for odd $p$ (or $[M,\rho]$ for even $p$) is written as a linear combination of $(EQ^\ell)\otimes(EQ^{\ell'})$ with $\ell,\ell'\ge 0$ and $\ell+\ell'\le p-1$.
Therefore, one can also calculate $\tr\rho^n$ for general odd (or even) $p$ with the same technique as $p=1$ (or $p=2$), by introducing
\begin{align}
\phi_m^{(\ell)}(Q)=\frac{(\rho^m\cdot EQ^\ell)(Q)}{E(Q)},
\end{align}
with $\ell=0,1,\cdots, p-1$.

\section{Application to models}\label{22&21}
In the previous section, we have introduced a systematic way to calculate the exact values of the partition function $Z(N)$ for the $(q,p)_k$ models.
In this section, we apply the method to the two cases, the $(2,2)_k$ model and the $(2,1)_k$ model, and give interpretations to the obtained exact results.
As we will see below, we observe that these models share some common properties with the ABJM matrix model which played important roles in the determination of the exact instanton expansion.
These properties again enable us to determine the instanton expansion in the $(2,2)_k$ model and the $(2,1)_k$ model.
We also observe an unexpected relation between these two models.

\subsection{$(2,2)_k$ model}
Applying the technique in section \ref{q2tba} to the $(2,2)_k$ model, we have computed the exact values of the partition function $Z_k(N)$ up to $N=N_\text{max}$ for $(k,N_\text{max})=(1,15),(2,13),(3,6),(4,7),(6,6)$.
The first few values are listed in table \ref{values}.
\begin{table}[!ht]
\begin{center}
\framebox{
\begin{minipage}{0.9\textwidth}
\begin{align*}
&Z_1(1)=\frac{1}{4\pi^2},\quad
Z_1(2)=\frac{15-\pi^2}{576\pi^4},\quad
Z_1(3)=\frac{855+75\pi^2-16\pi^4}{518400\pi^6},
\nonumber\\
&Z_2(1)=\frac{1}{8\pi^2},\quad
Z_2(2)=\frac{528-136\pi^2+9\pi^4}{73728\pi^4},\nonumber\\
&\quad Z_2(3)=\frac{67680-31200\pi^2+22454\pi^4-2025\pi^6}{265420800\pi^6},
\nonumber\\
&Z_3(1)=\frac{1}{12\pi^2},\quad
Z_3(2)=\frac{4131-1593\pi^2-128\sqrt{3}\pi^3+192\pi^4}{1259712\pi^4},
\nonumber\\&\quad
Z_3(3)=(22537035-19628325\pi^2-1296000\sqrt{3}\pi^3+15828048\pi^4
\nonumber\\&\qquad\qquad
+2188800\sqrt{3}\pi^5-2560000\pi^6)/(275499014400\pi^6),
\nonumber\\
&Z_4(1)=\frac{1}{16\pi^2},\quad
Z_4(2)=\frac{552-272\pi^2-72\pi^3+45\pi^4}{294912\pi^4},
\nonumber\\&\quad
Z_4(3)=\frac{152640-184800\pi^2-43200\pi^3+167482\pi^4+77400\pi^5
-38475\pi^6}{4246732800\pi^6},
\nonumber\\
&Z_6(1)=\frac{1}{24\pi^2},\quad
Z_6(2)=\frac{136080-92232\pi^2-25088\sqrt{3}\pi^3+21801\pi^4}
{161243136\pi^4},
\nonumber\\&\quad
Z_6(3)=(1565192160-2799360000\pi^2
-711244800\sqrt{3}\pi^3+2988770238\pi^4
\nonumber\\&\qquad\qquad
+1550649600\sqrt{3}\pi^5-1090902475\pi^6)
/(141055495372800\pi^6).
\end{align*}
\end{minipage}
}
\end{center}
\caption{Exact values of the partition function $Z_k(N)$ for the $(2,2)_k$ model.}
\label{values}
\end{table}
To obtain the non-perturbative part of the grand potential $J^{\text{np}}(\mu)$, we fit these data by the inverse transformation of \eqref{Jdef}
\begin{align}
Z(N)=\int_{-\infty}^\infty\frac{d\mu}{2\pi i}e^{J(\mu)-\mu N},
\end{align}
where the grand potential consists of the perturbative and non-perturbative parts
\begin{align}
J(\mu)=J^{\text{pert}}(\mu)+J^{\text{np}}(\mu).
\label{22fitZ}
\end{align}
The perturbative part $J^{\text{pert}}(\mu)$ is given as \eqref{Jpert}, where for the $(q,p)_k$ model the coefficients $C$ \cite{GHP,HKPT,MP} and $B$ \cite{MN1} are
\begin{align}
C=\frac{2}{\pi^2kqp},\quad
B=-\frac{1}{6k}\biggl[\frac{p}{q}+\frac{q}{p}-\frac{4}{qp}\biggr]
+\frac{k}{24}qp,
\label{CBqp}
\end{align}
and for $A$ we adopt the conjectural relation to that of the ABJM matrix model \cite{MN1}
\begin{align}
A=\frac{1}{2}(p^2A_\text{ABJM}(qk)+q^2A_\text{ABJM}(pk)).
\label{Aqpconj}
\end{align}
After subtracting the perturbative part, we can proceed to determine the instanton expansion of the non-perturbative part, as in \cite{HMO2} for the ABJM case.
The result is given in table \ref{instanton}.
\begin{table}[!ht]
\begin{center}
\framebox{
\begin{minipage}{0.9\textwidth}
\begin{align*}
J^\text{np}_{k=1}&=\frac{4\mu^2+4\mu+4}{\pi^2}e^{-\mu}
+\biggl[-\frac{26\mu^2+\mu+9/2}{\pi^2}+2\biggr]e^{-2\mu}
\nonumber\\&\quad
+\biggl[\frac{736\mu^2-608\mu/3+616/9}{3\pi^2}-32\biggr]e^{-3\mu}
\nonumber\\&\quad
+\biggl[-\frac{2701\mu^2-13949\mu/12+11291/48}{\pi^2}
+466\biggr]e^{-4\mu}
\nonumber\\&\quad
+\biggl[\frac{161824\mu^2-1268488\mu/15+1141012/75}{5\pi^2}
-6720\biggr]e^{-5\mu}
\nonumber\\&\quad
+\biggl[-\frac{1227440\mu^2-10746088\mu/15+631257/5}{3\pi^2}
+\frac{292064}{3}\biggr]e^{-6\mu}
\nonumber\\&\quad
+\biggl[\frac{37567744\mu^2-2473510336\mu/105+9211252832/2205}{7\pi^2}
-1420800\biggr]e^{-7\mu}
\nonumber\\&\quad
+{\mathcal O}(e^{-8\mu}),
\nonumber\\
J^\text{np}_{k=2}&=4e^{-\frac{1}{2}\mu}
+\biggl[\frac{2\mu^2+2\mu+2}{\pi^2}-6\biggr]e^{-\mu}
+\frac{16}{3}e^{-\frac{3}{2}\mu}
+\biggl[-\frac{13\mu^2+\mu/2+9/4}{\pi^2}-14\biggr]e^{-2\mu}
\nonumber\\&\quad
+\frac{544}{5}e^{-\frac{5}{2}\mu}
+\biggl[\frac{368\mu^2-304\mu/3+308/9}{3\pi^2}-288\biggr]e^{-3\mu}
-\frac{640}{7}e^{-\frac{7}{2}\mu}+{\mathcal O}(e^{-4\mu}),
\nonumber\\
J^\text{np}_{k=3}&=\frac{16}{3}e^{-\frac{1}{3}\mu}-4e^{-\frac{2}{3}\mu}
+\biggl[\frac{4\mu^2+4\mu+4}{3\pi^2}+\frac{128}{9}\biggr]e^{-\mu}
-\frac{613}{9}e^{-\frac{4}{3}\mu}+\frac{3536}{15}e^{-\frac{5}{3}\mu}
\nonumber\\&\quad
+\biggl[-\frac{26\mu^2+\mu+9/2}{3\pi^2}-\frac{7318}{9}\biggr]e^{2\mu}
+\frac{544352}{189}e^{-\frac{7}{3}\mu}+{\mathcal O}(e^{-\frac{8}{3}\mu}),
\nonumber\\
J^\text{np}_{k=4}&=8e^{-\frac{1}{4}\mu}
-8e^{-\frac{1}{2}\mu}
+\frac{80}{3}e^{-\frac{3}{4}\mu}
+\biggl[\frac{\mu^2+\mu+1}{\pi^2}-96\biggr]e^{-\mu}
+\frac{1888}{5}e^{-\frac{5}{4}\mu}
-\frac{4736}{3}e^{-\frac{3}{2}\mu}
\nonumber\\&\quad
+\frac{44416}{7}e^{-\frac{7}{4}\mu}+{\mathcal O}(e^{-2\mu}),
\nonumber\\
J^\text{np}_{k=6}&=16e^{-\frac{1}{6}\mu}-\frac{52}{3}e^{-\frac{1}{3}\mu}
+\frac{148}{3}e^{-\frac{1}{2}\mu}-189e^{-\frac{2}{3}\mu}
+\frac{4336}{5}e^{-\frac{5}{6}\mu}
\nonumber\\&\quad
+\biggl[\frac{2\mu^2+2\mu+2}{3\pi^2}-\frac{38102}{9}\biggr]e^{-\mu}
+\frac{446032}{21}e^{-\frac{7}{6}\mu}+{\mathcal O}(e^{-\frac{4}{3}\mu}).
\nonumber
\end{align*}
\end{minipage}
}
\end{center}
\caption{Instanton expansion in the $(2,2)_k$ model found by fitting to the exact values of the partition function in table \ref{values}.}
\label{instanton}
\end{table}

From the results in table \ref{instanton}, we expect that the worldsheet instanton exponent is given by $e^{-\frac{\mu}{k}}$ while the membrane instanton exponent is $e^{-\mu}$.
Together with their bound states, the non-perturbative part should be
\begin{align}
J^\text{np}(\mu)=\sum_{(\ell,m)\ne(0,0)}
f_{\ell,m}(\mu)e^{-(\ell+\frac{m}{k})\mu},
\label{flm}
\end{align}
where the pure membrane instanton and the pure worldsheet instanton are given by
\begin{align}
f_{\ell,0}(\mu)=a_\ell\mu^2+b_\ell\mu+c_\ell,\quad f_{0,m}=d_m.
\label{22mbabc}
\end{align}
For later convenience we introduce the following functions,
\begin{align}
J_a=\sum_{\ell=1}^\infty a_\ell e^{-\ell\mu},\quad
J_b=\sum_{\ell=1}^\infty b_\ell e^{-\ell\mu},\quad
J_c=\sum_{\ell=1}^\infty c_\ell e^{-\ell\mu}.
\end{align}

\subsubsection{Membrane instanton}
First we consider the pure membrane instantons.
As demonstrated in \cite{MN1}, the small $k$ expansion of the grand potential in the $(q,p)_k$ model can be systematically calculated by the method of the WKB expansion \cite{MP}.
We have calculated the grand potential up to ${\cal O}(k^9)$.
Then, expanding the results around $\mu\rightarrow\infty$ by using the formula in \cite{MN2},
we obtain, besides the perturbative part which is consistent with \eqref{CBqp} and \eqref{Aqpconj}, the explicit small $k$ expansion of the coefficients $a_\ell$, $b_\ell$ and $c_\ell$ in \eqref{22mbabc}.

For $a_\ell$, we find
\begin{align}
a_1&=\frac{2}{\pi^2k}+{\mathcal O}(k^9),\nonumber\\
a_2&=-\frac{9}{\pi^2k}+2k-\frac{2\pi^2k^3}{3}
+\frac{4\pi^4k^5}{45}-\frac{2\pi^6k^7}{315}+{\mathcal O}(k^9),
\nonumber\\
a_3&=\frac{200}{3\pi^2k}-32k+\frac{32\pi^2k^3}{3}-\frac{64\pi^4k^5}{45}
+\frac{32\pi^6k^7}{315}+{\mathcal O}(k^9),
\nonumber\\
a_4&=-\frac{1225}{2\pi^2k}+500k-\frac{752\pi^2k^3}{3}
+\frac{704\pi^4k^5}{9}-\frac{5792\pi^6k^7}{315}+{\mathcal O}(k^9),
\nonumber\\
a_5&=\frac{31752}{5\pi^2k}-7840k+\frac{17440\pi^2k^3}{3}
-\frac{26944\pi^4k^5}{9}+\frac{10592\pi^6k^7}{9}+{\mathcal O}(k^9).
\end{align}
From these results we can find directly the following expressions for finite $k$,
\begin{align}
a_1&=\frac{2}{\pi^2k},\nonumber\\
a_2&=-\frac{8+\cos 2\pi k}{\pi^2k},\nonumber\\
a_3&=\frac{152+48\cos 2\pi k}{3\pi^2k},\nonumber\\
a_4&=-\frac{788+416\cos 2\pi k+21\cos 4\pi k}{2\pi^2k},\nonumber\\
a_5&=\frac{17352+12800\cos 2\pi k+1520\cos 4\pi k+80\cos 6\pi k}
{5\pi^2k}.
\label{a22}
\end{align}
Our criterion to decide the ansatz of the expression is that the function forms should be similar to the ABJM case and should not be too complicated.
Of course, it is reasonable to doubt whether we can determine the entire functions just from the first five terms of the series expansions.
However, we will continue this kind of arguments from now on and provide non-trivial checks to the results later from time to time.

Before going on to conjecture the explicit form of the remaining part of the instanton coefficients, we introduce the effective chemical potential
\begin{align}
\mu_\text{eff}=\mu+\frac{J_a}{C},
\label{mueffofmu}
\end{align}
in terms of which the quadratic part of the instanton coefficients are absorbed into the perturbative part as
\begin{align}
\frac{C}{3}\mu^3+B\mu+A+J_a\mu^2+J_b\mu+J_c=\frac{C}{3}\mu_\text{eff}^3+B\mu_\text{eff}+A+{\widetilde J}_b\mu_\text{eff}+{\widetilde J}_c,
\end{align}
and call the instanton coefficients in ${\widetilde J}_b$ and ${\widetilde J}_c$ as ${\widetilde b}_\ell$ and ${\widetilde c}_\ell$ respectively.
Then we find that these instanton coefficients satisfy the following derivative relation (at least up to ${\cal O}(k^9)$)
\begin{align}
\widetilde c_\ell=-k^2\frac{d}{dk}\frac{\widetilde b_\ell}{\ell k}.
\label{22der}
\end{align}
Note that the same reduction of instanton coefficients also occurred in the ABJM matrix model \cite{HMO3} and the $(2,1)_k$ model \cite{MN1}.
In the ABJM case, the effective chemical potential played an important role to handle the bound states of the worldsheet instantons and the membrane instantons \cite{HMO3}, and the derivative relation was crucial to write down the explicit formula for all order membrane instanton corrections as in \eqref{Jtop} \cite{HMMO}.
As we will see below, they play just the same role in the $(2,2)_k$ model.

The first few coefficients $\widetilde b_\ell$ are given by
\begin{align}
&\widetilde b_1=-\frac{4}{\pi^2k}+\frac{4k}{3}+\frac{4\pi^2k^3}{45}
+\frac{8\pi^4k^5}{945}+\frac{4\pi^6k^7}{4725}+{\mathcal O}(k^9),
\nonumber\\
&\widetilde b_2=\frac{9}{\pi^2k}-6k+\frac{14\pi^2k^3}{5}
-\frac{44\pi^4k^5}{105}+\frac{18\pi^6k^7}{175}+{\mathcal O}(k^9),
\nonumber\\
&\widetilde b_3=-\frac{328}{9\pi^2k}+\frac{184k}{3}
-\frac{152\pi^2k^3}{5}+\frac{752\pi^4k^5}{105}+\frac{8\pi^6k^7}{525}
+{\mathcal O}(k^9),
\nonumber\\
&\widetilde b_4=\frac{777}{4\pi^2k}-598k+\frac{9704\pi^2k^3}{15}
-\frac{16448\pi^4k^5}{45}+\frac{202304\pi^6k^7}{1575}
+{\mathcal O}(k^9),
\nonumber\\
&\widetilde b_5=-\frac{30004}{25\pi^2k}+\frac{18004k}{3}
-\frac{96700\pi^2k^3}{9}+\frac{1957000\pi^4k^5}{189}
-\frac{169748\pi^6k^7}{27}
+{\mathcal O}(k^9).
\end{align}
From them, we can read off
\begin{align}
&\widetilde b_1=\frac{-4\cos\pi k}{\pi\sin\pi k},\nonumber\\
&\widetilde b_2=\frac{9+8\cos 2\pi k+\cos 4\pi k}{\pi\sin 2\pi k},
\nonumber\\
&\widetilde b_3=\frac{-4(45\cos\pi k+28\cos 3\pi k+9\cos 5\pi k)}
{3\pi\sin 3\pi k}.
\end{align}
Interestingly, we find that these coefficients satisfy the following multi-covering structure,
\begin{align}
\widetilde b_1&=\beta_1(k),&
\beta_1(k)&=-\frac{2\sin 2\pi k}{\pi\sin^2\pi k},\nonumber\\
\widetilde b_2&=\frac{1}{2}\beta_1(2k)+\beta_2(k),&
\beta_2(k)&=\frac{8\sin 2\pi k+\sin 4\pi k}{2\pi\sin^2\pi k},
\nonumber\\
\widetilde b_3&=\frac{1}{3}\beta_1(3k)+\beta_3(k),&
\beta_3(k)&=-\frac{6\sin 2\pi k+6\sin 4\pi k}{\pi\sin^2\pi k}.
\label{b22123}
\end{align}
If we further assume the multi-covering structure,
\begin{align}
\widetilde b_4&=\frac{1}{4}\beta_1(4k)
+\frac{1}{2}\beta_2(2k)+\beta_4(k),\nonumber\\
\widetilde b_5&=\frac{1}{5}\beta_1(5k)+\beta_5(k),
\label{b2245}
\end{align}
we can proceed to determine the $k$ dependence for higher instantons as
\begin{align}
\beta_4(k)
&=\frac{9\sin 2\pi k+30\sin 4\pi k+9\sin 6\pi k}{\pi\sin^2\pi k},
\nonumber\\
\beta_5(k)&=-\frac{20\sin 2\pi k+100\sin 4\pi k
+100\sin 6\pi k+20\sin 8\pi k}{\pi\sin^2\pi k}.
\label{b22beta}
\end{align}

\subsubsection{Effective chemical potential}
In \eqref{a22} we have found that the series expansions of $a_\ell$ match well with the ansatz that the argument of the cosine functions in the numerator of $a_\ell$ is always a multiple of $2\pi k$.
As in the ABJM case \cite{HMO3}, if we assume that this is true for all instantons, due to the periodicity of the cosine function, we find that, when $k$ is an integer, the value of the function $a_\ell$ is the same as its leading term in the WKB expansion,
\begin{align}
a_\ell=\frac{(ka_\ell)\big|_{k=0}}{k}.
\end{align}
The expression of the leading term in the small $k$ expansion was known for all instantons \cite{MN2}.
Picking up the coefficients of the $\mu^2$ terms, for the $(q,p)_k$ model we found
\begin{align}
(kJ_a)\big|_{k=0}
&=-\frac{1}{\pi^2qp}
\sum_{r\in{\mathbb N}/\gcd(q,p,2)}\frac{1}{r\cos 2\pi r}
\frac{\Gamma(2qr+1)}{\Gamma(qr+1)^2}\frac{\Gamma(2pr+1)}{\Gamma(pr+1)^2}
e^{-2r\mu}.
\label{Ja}
\end{align}
After substituting $(q,p)=(2,2)$ we find
\begin{align}
J_a=\frac{2e^{-\mu}}{\pi^2k}
{}_4F_3\Bigl(1,1,\frac{3}{2},\frac{3}{2};2,2,2;-16e^{-\mu}\Bigr).
\end{align}
Hence, at integral $k$, the expression relating the effective chemical potential and the original one is
\begin{align}
\mu_\text{eff}=\mu+4e^{-\mu}
{}_4F_3\Bigl(1,1,\frac{3}{2},\frac{3}{2};2,2,2;-16e^{-\mu}\Bigr).
\label{mueff22}
\end{align}
If we express the instanton expansion in table \ref{instanton} using this effective chemical potential, the coefficients of $\pi^{-2}$ become somewhat simpler,
\begin{align}
J^\text{np}_k=\sum_{\ell=1}^\infty\frac{f_{k,\ell}}{\pi^2}
\biggl(\frac{\ell^2}{2}\mu_\text{eff}^2+\ell\mu_\text{eff}+1\biggr)
e^{-\ell\mu_\text{eff}}
+\sum_{m=1}^\infty g_{k,m}e^{-m\frac{\mu_\text{eff}}{k}},
\label{mupoly}
\end{align}
with rational numbers $f_{k,\ell}$ and $g_{k,m}$.
See table \ref{mueff}.
\begin{table}[!ht]
\begin{center}
\framebox{
\begin{minipage}{0.9\textwidth}
\begin{align*}
J^\text{np}_{k=1}&=\frac{2(\mu_\text{eff}^2+2\mu_\text{eff}+2)}{\pi^2}
e^{-\mu_\text{eff}}
+\biggl[-\frac{9(2\mu_\text{eff}^2+2\mu_\text{eff}+1)}{2\pi^2}
+2\biggr]e^{-2\mu_\text{eff}}
\nonumber\\&\quad
+\biggl[\frac{164(9\mu_\text{eff}^2+6\mu_\text{eff}+2)}{27\pi^2}
-16\biggr]e^{-3\mu_\text{eff}}
\nonumber\\&\quad
+\biggl[-\frac{777(8\mu_\text{eff}^2+4\mu_\text{eff}+1)}{16\pi^2}
+138\biggr]e^{-4\mu_\text{eff}} 
\nonumber\\&\quad
+\biggl[\frac{15002(25\mu_\text{eff}^2+10\mu_\text{eff}+2)}{125\pi^2}
-1216\biggr]e^{-5\mu_\text{eff}}
\nonumber\\&\quad
+\biggl[-\frac{4073(18\mu_\text{eff}^2+6\mu_\text{eff}+1)}{3\pi^2}
+\frac{32852}{3}\biggr]e^{-6\mu_\text{eff}}
\nonumber\\&\quad
+\biggl[\frac{1445404(49\mu_\text{eff}^2+14\mu_\text{eff}+2)}{343\pi^2}
-100272\biggr]e^{-7\mu_\text{eff}}
+{\mathcal O}(e^{-8\mu_\text{eff}}),
\nonumber\\
J^\text{np}_{k=2}&=4e^{-\frac{1}{2}\mu_\text{eff}}
+\biggl[\frac{\mu_\text{eff}^2+2\mu_\text{eff}+2}{\pi^2}
-7\biggr]e^{-\mu_\text{eff}}
+\frac{40}{3}e^{-\frac{3}{2}\mu_\text{eff}}
\nonumber\\&\quad
+\biggl[-\frac{9(2\mu_\text{eff}^2+2\mu_\text{eff}+1)}{4\pi^2}
-\frac{75}{2}\biggr]e^{-2\mu_\text{eff}}
+\frac{724}{5}e^{-\frac{5}{2}\mu_\text{eff}}
\nonumber\\&\quad
+\biggl[\frac{82(9\mu_\text{eff}^2+6\mu_\text{eff}+2)}{27\pi^2}
-\frac{1318}{3}\biggr]e^{-3\mu_\text{eff}}
+\frac{7704}{7}e^{-\frac{7}{2}\mu_\text{eff}}
+{\mathcal O}(e^{-4\mu_\text{eff}}),
\nonumber\\
J^\text{np}_{k=3}&=\frac{16}{3}e^{-\frac{1}{3}\mu_\text{eff}}
-4e^{-\frac{2}{3}\mu_\text{eff}}
+\biggl[\frac{2(\mu_\text{eff}^2+2\mu_\text{eff}+2)}{3\pi^2}
+\frac{112}{9}\biggr]e^{-\mu_\text{eff}}
-61e^{-\frac{4}{3}\mu_\text{eff}}
\nonumber\\&\quad
+\frac{3376}{15}e^{-\frac{5}{3}\mu_\text{eff}}
+\biggl[-\frac{3(2\mu_\text{eff}^2+2\mu_\text{eff}+1)}{2\pi^2}
-\frac{2266}{3}\biggr]e^{-2\mu_\text{eff}}
+\frac{52880}{21}e^{-\frac{7}{3}\mu_\text{eff}}
\nonumber\\&\quad
+{\mathcal O}(e^{-\frac{8}{3}\mu_\text{eff}}),
\nonumber\\
J^\text{np}_{k=4}&=8e^{-\frac{1}{4}\mu_\text{eff}}
-8e^{-\frac{1}{2}\mu_\text{eff}}
+\frac{80}{3}e^{-\frac{3}{4}\mu_\text{eff}}
+\biggl[\frac{\mu_\text{eff}^2+2\mu_\text{eff}+2}{2\pi^2}
-\frac{197}{2}\biggr]e^{-\mu_\text{eff}}
+\frac{1928}{5}e^{-\frac{5}{4}\mu_\text{eff}}
\nonumber\\&\quad
-\frac{4784}{3}e^{-\frac{3}{2}\mu_\text{eff}}
+\frac{44976}{7}e^{-\frac{7}{4}\mu_\text{eff}}
+{\mathcal O}(e^{-2\mu_\text{eff}}),
\nonumber\\
J^\text{np}_{k=6}&=16e^{-\frac{1}{6}\mu_\text{eff}}
-\frac{52}{3}e^{-\frac{1}{3}\mu_\text{eff}}
+\frac{148}{3}e^{-\frac{1}{2}\mu_\text{eff}}
-189e^{-\frac{2}{3}\mu_\text{eff}}
+\frac{4336}{5}e^{-\frac{5}{6}\mu_\text{eff}}
\nonumber\\&\quad
+\biggl[\frac{\mu_\text{eff}^2+2\mu_\text{eff}+2}{3\pi^2}
-\frac{38137}{9}\biggr]e^{-\mu_\text{eff}}
+\frac{148752}{7}e^{-\frac{7}{6}\mu_\text{eff}}
+{\mathcal O}(e^{-\frac{4}{3}\mu_\text{eff}}).
\end{align*}
\end{minipage}
}
\end{center}
\caption{Instanton expansion in the $(2,2)_k$ model in table \ref{instanton} rewritten in terms of the effective chemical potential $\mu_\text{eff}$.}
\label{mueff}
\end{table}

\subsubsection{Worldsheet instanton}\label{22ws}
Now let us guess the $k$ dependence of the worldsheet instanton coefficients by looking at table \ref{mueff} more carefully.
As the membrane instanton coefficients $\widetilde b_\ell$ we have guessed above \eqref{b22123} and \eqref{b22beta} diverge when $\ell k\in\mathbb{Z}$, we expect that the worldsheet instanton coefficients $d_m$ also diverge when $m/k\in\mathbb{Z}$ so that the total non-perturbative effects are finite after the cancellation of divergences, as in the ABJM case \cite{HMO2,CM}.
From this fact and the experience in the ABJM case \cite{HMO2}, we expect that $d_m$ is expressed as a linear combination of $\bigl(\sin\frac{m\pi}{dk}\bigr)^{2g-2}$ with $d$ being a divisor of $m$ and $g$ being the genus.
Then using the coefficients of $e^{-m\frac{\mu_\text{eff}}{k}}$ at $k=2,3,4,6$ for $m=1$, those at $k=3,4,6$ for $m=2$ and those at $k=4,6$ for $m=3$ (namely, those at $k\ge m$), we find the first few worldsheet instantons are given by
\begin{align}
d_1=\frac{4}{\sin^2\frac{\pi}{k}},\quad
d_2=\frac{2}{\sin^2\frac{2\pi}{k}}-\frac{5}{\sin^2\frac{\pi}{k}},\quad
d_3=\frac{4}{3\sin^2\frac{3\pi}{k}}+\frac{12}{\sin^2\frac{\pi}{k}}.
\label{22d123}
\end{align}
Interestingly, we have observed that several properties of the ABJM matrix model also hold here.
\begin{itemize}
\item[$\circ$]
The result \eqref{22d123} also matches with the coefficient at $e^{-\frac{3}{2}\mu_\text{eff}}$ and $k=2$ in table \ref{mueff}, where there could exist contributions from the bound state.
This means that our rewriting with the effective chemical potential $\mu_\text{eff}$ automatically takes care of the bound states.
\item[$\circ$]
Although the coefficients of both the worldsheet instanton and the membrane instanton are divergent at $e^{-\mu_\text{eff}}$ and $k=1,2,3$, at $e^{-2\mu_\text{eff}}$ and $k=1,2$, and also at $e^{-3\mu_\text{eff}}$ and $k=1$, these divergences are completely cancelled and the finite results from the numerical fitting in table \ref{mueff} are reproduced.
\item[$\circ$]
Finally, the first terms in these coefficients
\begin{align}
d_m\simeq\frac{4}{m\sin^2\frac{m\pi}{k}}
\end{align}
correctly recover the multi-covering structure expected for the topological string theory,
\begin{align}
d_1&=\delta_1(k),&\delta_1(k)&=\frac{4}{\sin^2\frac{\pi}{k}},
\nonumber\\
d_2&=\frac{1}{2}\delta_1\Bigl(\frac{k}{2}\Bigr)+\delta_2(k),&
\delta_2(k)&=-\frac{5}{\sin^2\frac{\pi}{k}},
\nonumber\\
d_3&=\frac{1}{3}\delta_1\Bigl(\frac{k}{3}\Bigr)+\delta_3(k),&
\delta_3(k)&=\frac{12}{\sin^2\frac{\pi}{k}}.
\label{d22123}
\end{align}
\end{itemize}

After observing that the bound states are correctly taken care of by the effective chemical potential and that the expression has the multi-covering structure, we have more equations and less unknowns for the higher worldsheet instantons.
Using the remaining data in table \ref{mueff}, we can further find
\begin{align}
d_4
&=\frac{1}{4}\delta_1\Bigl(\frac{k}{4}\Bigr)
+\frac{1}{2}\delta_2\Bigl(\frac{k}{2}\Bigr)
+\delta_4(k),&
\delta_4(k)&=-\frac{48}{\sin^2\frac{\pi}{k}}+5,
\nonumber\\
d_5
&=\frac{1}{5}\delta_1\Bigl(\frac{k}{5}\Bigr)
+\delta_5(k),&
\delta_5(k)&=\frac{240}{\sin^2\frac{\pi}{k}}-96.
\label{d2245}
\end{align}

\subsubsection{Topological string theory}\label{22tst}

From the fact that the non-perturbative effects in the $(2,2)_k$ model share many structures found in the ABJM matrix model, we expect that they can also be described using the free energy of the refined topological string theory as in \eqref{Jtop}.
In the following, we find that this is actually the case.

By comparing the exponents of the worldsheet and membrane instantons, we can tentatively identify the K\"{a}hler parameter and the string coupling as
\begin{align}
T_\text{eff}=\frac{\mu_\text{eff}}{k},\quad g_s=\frac{1}{k}.
\end{align}
Note that although in the K\"{a}hler parameters of the ABJM case we have imaginary contributions coming from the discrete $B$ field \cite{DMP1}, we expect that in the current case the imaginary contributions are effectively absent because there are no signs $(-1)^{nd}$ appearing in the multi-covering structure \eqref{b22123}, \eqref{b2245}, \eqref{d22123} and \eqref{d2245}.
Combining the results for $\widetilde b_\ell$ and $d_m$ in \eqref{b22123}, \eqref{b22beta}, \eqref{d22123}, \eqref{d2245}, we can consistently identify the BPS indices $N^{\bf d}_{j_L,j_R}$ as in table \ref{BPS}.
Slightly differently, in the identification we encounter overall signs $(-1)^{d-1}$ for the BPS indices.
We have dropped these overall signs in table \ref{BPS} from the expectation that the BPS index should be non-negative.

Note that there are some ambiguities in determining the BPS indices $N^{\bf d}_{j_L,j_R}$ for $d=4,5$, as in table \ref{BPS}.
In spite of this, the diagonal Gopakumar-Vafa invariants \eqref{NtoGVinv}, which do not distinguish one of the spins, can be read off directly from the expression of the worldsheet instantons by
\begin{align}
\delta_d(k)=\sum_{g=0}^\infty n_g^d\biggl(2\sin\frac{\pi}{k}\biggr)^{2g-2}.
\end{align}
See table \ref{GV} ($1\le d\le 5$).
Surprisingly, we have observed the following property.
\begin{itemize}
\item
The diagonal Gopakumar-Vafa invariants of the $(2,2)_k$ model match with those of the local $D_5$ del Pezzo geometry \cite{KKV}, although the BPS indices \cite{HKP} are different.
\end{itemize}

\subsubsection{Quantum mirror map}
In the analysis so far, we have basically concentrated on the expression after introducing the effective chemical potential $\mu_\text{eff}$.
Lastly, for the $(2,2)_k$ model, let us comment on some interesting structures on the coefficients $a_\ell$.

In \cite{HMO3} it was found that, for the ABJM matrix model, the instanton coefficients $e_\ell$ appearing when we express $\mu$ in terms of $\mu_\text{eff}$ are somewhat simpler than the original ones $a_\ell$ appearing when we express $\mu_\text{eff}$ in terms of $\mu$.
Defining the same quantity for the $(2,2)_k$ model,
\begin{align}
\mu=\mu_\text{eff}+\frac{1}{C}\sum_{\ell=1}^\infty
e_\ell e^{-\ell\mu_\text{eff}},
\end{align}
we obtain
\begin{align}
e_1&=-4,\nonumber\\
e_2&=2\cos 2\pi k,\nonumber\\
e_3&=-\frac{8}{3}(2+3\cos 2\pi k),\nonumber\\
e_4&=16+32\cos 2\pi k+17\sin 4\pi k,\nonumber\\
e_5&=-\frac{4}{5}(101+200\cos 2\pi k+160\cos 4\pi k+40\cos 6\pi k),
\end{align}
which look simpler than $a_\ell$ in \eqref{a22}.
More interestingly, they also have the following multi-covering structure:
\begin{align}
e_1&=\epsilon_1(k),&\epsilon_1(k)&=-4,\nonumber\\
e_2&=\frac{1}{2}\epsilon_1(2k)+\epsilon_2(k),
&\epsilon_2(k)&=2(1+\cos 2\pi k),\nonumber\\
e_3&=\frac{1}{3}\epsilon_1(3k)+\epsilon_3(k),
&\epsilon_3(k)&=-4(1+2\cos 2\pi k),\nonumber\\
e_4&=\frac{1}{4}\epsilon_1(4k)+\frac{1}{2}\epsilon_2(2k)+\epsilon_4(k),
&\epsilon_4(k)&=16(1+2\cos 2\pi k+\cos 4\pi k),\nonumber\\
e_5&=\frac{1}{5}\epsilon_1(5k)+\epsilon_5(k),
&\epsilon_5(k)&=-16(5+10\cos 2\pi k+8\cos 4\pi k+2\cos 6\pi k),
\end{align}
as in the ABJM case \cite{HMMO}.
The integrality of the coefficients in the last expression would imply that they can be interpreted as topological invariants associated to the quantum mirror map.

\subsection{$(2,1)_k$ model}

\begin{table}[!ht]
\begin{center}
\framebox{
\begin{minipage}{0.9\textwidth}
\begin{align*}
J^\text{np}_{k=1}&=-\frac{4\mu^2+2\mu+1}{\pi^2}e^{-2\mu}
+\biggl[-\frac{26\mu^2+\mu/2+9/8}{\pi^2}+2\biggr]e^{-4\mu}
\nonumber\\
&\quad+\biggl[-\frac{736\mu^2-304\mu/3+154/9}{3\pi^2}+32\biggr]e^{-6\mu}
\nonumber\\
&\quad+\biggl[-\frac{2701\mu^2-13949\mu/24+11291/192}{\pi^2}
+466\biggr]e^{-8\mu}
+{\mathcal O}(e^{-10\mu}),
\\
J^\text{np}_{k=2}&=\frac{2\mu+2}{\pi}e^{-\mu}
+\biggl[-\frac{10\mu^2+7\mu+7/2}{\pi^2}+1\biggr]e^{-2\mu}
+\frac{88\mu+52/3}{3\pi}e^{-3\mu}
\nonumber\\
&\quad+\biggl[-\frac{269\mu^2+193\mu/4+265/16}{\pi^2}+58\biggr]e^{-4\mu}
+\frac{4792\mu+1102/5}{5\pi}e^{-5\mu}
\nonumber\\&\quad
+\biggl[-\frac{31024\mu^2+736\mu/3+6443/9}{3\pi^2}
+\frac{9088}{3}\biggr]e^{-6\mu}
+\frac{277408\mu-31656/7}{7\pi}e^{-7\mu}
\nonumber\\
&\quad+{\mathcal O}(e^{-8\mu}),
\\
J^\text{np}_{k=3}&=\frac{8}{3}e^{-\frac{2}{3}\mu}-6e^{-\frac{4}{3}\mu}
+\biggl[-\frac{4\mu^2+2\mu+1}{3\pi^2}+\frac{88}{9}\biggr]e^{-2\mu}
-\frac{238}{9}e^{-\frac{8}{3}\mu}+\frac{848}{15}e^{-\frac{10}{3}\mu}
\nonumber\\
&\quad+\biggl[-\frac{26\mu^2+\mu/2+9/8}{3\pi^2}
-\frac{1540}{9}\biggr]e^{-4\mu}
+\frac{82672}{189}e^{-\frac{14}{3}\mu}
+{\mathcal O}(e^{-\frac{16}{3}\mu}),
\\
J^\text{np}_{k=4}&=2\sqrt{2}e^{-\frac{1}{2}\mu}
+\biggl[\frac{\mu+1}{\pi}-4\biggr]e^{-\mu}
+\frac{16\sqrt{2}}{3}e^{-\frac{3}{2}\mu}
+\biggl[-\frac{10\mu^2+7\mu+7/2}{2\pi^2}-\frac{45}{2}\biggr]
e^{-2\mu}
\nonumber\\&
\quad+\frac{288\sqrt{2}}{5}e^{-\frac{5}{2}\mu}
+\biggl[\frac{44\mu+26/3}{3\pi}-\frac{640}{3}\biggr]e^{-3\mu}
+\frac{2816\sqrt{2}}{7}e^{-\frac{7}{2}\mu}
+{\mathcal O}(e^{-4\mu}),
\\
J^\text{np}_{k=6}&=\frac{8}{\sqrt{3}}e^{-\frac{1}{3}\mu}
-\frac{14}{3}e^{-\frac{2}{3}\mu}
+\biggl[\frac{2\mu+2}{3\pi}+\frac{24}{\sqrt{3}}\biggr]e^{-\mu}
-\frac{154}{3}e^{-\frac{4}{3}\mu}
+\frac{1472}{5\sqrt{3}}e^{-\frac{5}{3}\mu}
\nonumber\\
&\quad+\biggl[-\frac{10\mu^2+7\mu+7/2}{3\pi^2}
-\frac{4883}{9}\biggr]e^{-2\mu}
+\frac{20992}{7\sqrt{3}}e^{-\frac{7}{3}\mu}
+{\mathcal O}(e^{-\frac{8}{3}\mu}).
\end{align*}
\end{minipage}
}
\end{center}
\caption{Instanton expansion in the $(2,1)_k$ model found by fitting to the exact values of the partition function in \cite{MN1}.}
\label{21instanton}
\end{table}

In \cite{MN1} we took the first step to study the non-perturbative effects of the $(2,1)_k$ model.
Due to the lack of comparison with other models, we were not able to find a concrete structure at that time, except the pole cancellation mechanism in the first few instantons.
Now that we have looked at the $(2,2)_k$ model, let us revisit the $(2,1)_k$ model here.
The non-perturbative part of the grand potential is given in table \ref{21instanton}.
From this result we find that the instanton expansion takes a similar form as the $(2,2)_k$ model \eqref{flm}.
The difference is that the worldsheet instanton exponent is $e^{-2\mu/k}$ in the $(2,1)_k$ model (instead of $e^{-\mu/k}$ in the $(2,2)_k$ model) and that the membrane instanton expansion
\begin{align}
J^\text{MB}(\mu)=J_a\mu^2+J_b\mu+J_c,
\end{align}
takes different forms for the odd and even instanton numbers,
\begin{align}
J_a=\sum_{\ell=1}^\infty a_{2\ell}e^{-2\ell\mu},\quad
J_b=\sum_{\ell=1}^\infty b_{2\ell}e^{-2\ell\mu},\quad
J_c=\sum_{\ell=1}^\infty c_{\ell}e^{-\ell\mu},
\end{align}
as observed in \cite{MN1}.

\subsubsection{Membrane instanton}\label{21mb}
Again we start with the WKB expansion of the membrane instanton.
From the explicit computation up to ${\mathcal O}(k^9)$ as in the $(2,2)_k$ model we directly find
\begin{align}
a_2=-\frac{2(2+\cos\pi k)}{\pi^2k},\quad
a_4=-\frac{44+48\cos\pi k+13\cos 2\pi k}{\pi^2k}.
\label{a21}
\end{align}
Also, as already found in \cite{MN1}, if we introduce the effective chemical potential $\mu_\text{eff}$ as we have done in the $(2,2)_k$ model \eqref{mueffofmu}, the quadratic part is absorbed into the perturbative part, and the linear part and the constant part of the new instanton coefficients $\widetilde b_{2\ell}$ and $\widetilde c_{2\ell}$ satisfy the following derivative relation,
\begin{align}
\widetilde c_{2\ell}
=-k^2\frac{d}{dk}\frac{\widetilde b_{2\ell}}{2\ell k}.
\label{21der}
\end{align}

Using the results of the WKB expansion up to ${\cal O}(k^9)$, we find that the first few membrane instantons are given by
\begin{align}
\widetilde c_1=-\frac{2\cos\frac{\pi k}{2}}{\sin\frac{\pi k}{2}},
\quad
\widetilde b_2=\frac{5+8\cos\pi k+\cos 2\pi k}{\pi\sin\pi k},
\quad
\widetilde c_3
=-\frac{2(15\cos\frac{\pi k}{2}+8\cos\frac{3\pi k}{2}
+3\cos\frac{5\pi k}{2})}{3\sin\frac{3\pi k}{2}}.
\end{align}
After noticing the ``multi-covering'' structure, we can proceed to determine the coefficients of higher order instantons as
\begin{align}
\widetilde c_1&=\gamma_1(k),&
\gamma_1(k)&=-\frac{\sin\pi k}{\sin^2\frac{\pi k}{2}},
\nonumber\\
\widetilde b_2&=\frac{-1}{\pi}\gamma_1(2k)+\beta_2(k),&
\beta_2(k)&=\frac{4\sin\pi k+\sin 2\pi k}{2\pi\sin^2\frac{\pi k}{2}},
\nonumber\\
\widetilde c_3&=\frac{-1}{3}\gamma_1(3k)+\gamma_3(k),&
\gamma_3(k)&=-\frac{\sin\pi k+\sin 2\pi k}{\sin^2\frac{\pi k}{2}},
\nonumber\\
\widetilde b_4&=\frac{1}{2\pi}\gamma_1(4k)+\frac{1}{2}\beta_2(2k)
+\beta_4(k),&
\beta_4(k)&=\frac{16\sin\pi k+23\sin 2\pi k+16\sin 3\pi k+5\sin 4\pi k}
{\pi\sin^2\pi k},\nonumber\\
\widetilde c_5&=\frac{1}{5}\gamma_1(5k)+\gamma_5(k),&
\gamma_5(k)&=-\frac{2\sin\pi k+6\sin 2\pi k+6\sin 3\pi k+2\sin 4\pi k}
{\sin^2\frac{\pi k}{2}}.
\label{cb}
\end{align}
Note that the structure in \eqref{cb} is tentatively assumed to reduce the number of unknowns, though it is probably not the multi-covering structure compatible with the cancellation mechanism.
For example, relative signs may appear depending on the spins $(j_L,j_R)$.

\subsubsection{Effective chemical potential}
As in the $(2,2)_k$ model let us try to use the effective chemical potential also in the $(2,1)_k$ model.
Suppose again that the arguments of the cosine functions in the numerators of $a_\ell$ are multiples of $\pi k$ as in \eqref{a21}, then when $k$ is even, we have
\begin{align}
a_\ell=\frac{(ka_\ell)\big|_{k=0}}{k}.
\end{align}
Using \eqref{Ja} we find
\begin{align}
J_a=-\frac{6e^{-2\mu}}{\pi^2k}
{}_4F_3\Bigl(1,1,\frac{7}{4},\frac{5}{4};2,2,2;64e^{-2\mu}\Bigr),
\end{align}
which implies
\begin{align}
\mu_\text{eff}=\mu-6e^{-2\mu}
{}_4F_3\Bigl(1,1,\frac{7}{4},\frac{5}{4};2,2,2;64e^{-2\mu}\Bigr).
\end{align}
For odd $k$, we conjecture a relation
\begin{align}
\mu_\text{eff}=\mu-2e^{-2\mu}
{}_4F_3\Bigl(1,1,\frac{3}{2},\frac{3}{2};2,2,2;16e^{-2\mu}\Bigr),
\label{mu21odd}
\end{align}
similar to those for the ABJM matrix model \cite{HMO3} and the $(2,2)_k$ model \eqref{mueff22}.
Although we do not have a logical reason for \eqref{mu21odd}, this is motivated by the following observations: the $J^{\text{np}}_{k=1}$ in table \ref{21instanton} looks similar to the $J^{\text{np}}_{k=1}$ of the ABJM matrix model \cite{HMO2,HMO3} and the $(2,2)_k$ model in table \ref{instanton};
the relation \eqref{mu21odd} is consistent with $a_2$ and $a_4$ in \eqref{a21}; the relation \eqref{mu21odd} simplifies the expressions of the instanton expansion as in \eqref{mupoly}.
The grand potential in terms of the effective chemical potential is given in table \ref{21mueff}.
\begin{table}[!ht]
\begin{center}
\framebox{
\begin{minipage}{0.9\textwidth}
\begin{align*}
J^\text{np}_{k=1}
&=-\frac{2\mu_\text{eff}^2+2\mu_\text{eff}+1}{\pi^2}e^{-2\mu_\text{eff}}
+\biggl[-\frac{9(8\mu_\text{eff}^2+4\mu_\text{eff}+1)}{8\pi^2}
+2\biggr]e^{-4\mu_\text{eff}}
\nonumber\\&\quad
+\biggl[-\frac{82(18\mu_\text{eff}^2+6\mu_\text{eff}+1)}{27\pi^2}+16\biggr]e^{-6\mu_\text{eff}}
\nonumber\\&\quad
+\biggl[-\frac{777(32\mu_\text{eff}^2+8\mu_\text{eff}+1)}{64\pi^2}+138\biggr]e^{-8\mu_\text{eff}}
+{\mathcal O}(e^{-10\mu_\text{eff}}),
\nonumber\\
J^\text{np}_{k=2}&=\frac{2(\mu_\text{eff}+1)}{\pi}e^{-\mu_\text{eff}}
+\biggl[-\frac{7(2\mu_\text{eff}^2+2\mu_\text{eff}+1)}{2\pi^2}
+\frac{7}{4}\biggr]e^{-2\mu_\text{eff}}
+\frac{52(3\mu_\text{eff}+1)}{9\pi}e^{-3\mu_\text{eff}}
\nonumber\\&\quad
+\biggl[-\frac{265(8\mu_\text{eff}^2+4\mu_\text{eff}+1)}{16\pi^2}
+\frac{401}{8}\biggr]e^{-4\mu_\text{eff}}
+\frac{2002(5\mu_\text{eff}+1)}{25\pi}e^{-5\mu_\text{eff}}
\nonumber\\&\quad
+\biggl[-\frac{5471(18\mu_\text{eff}^2+6\mu_\text{eff}+1)}{27\pi^2}
+\frac{10307}{6}\biggr]e^{-6\mu_\text{eff}}
+\frac{83004(7\mu_\text{eff}+1)}{49\pi}e^{-7\mu_\text{eff}}
\nonumber\\&\quad
+{\mathcal O}(e^{-8\mu_\text{eff}}),
\nonumber\\
J^\text{np}_{k=3}&=\frac{8}{3}e^{-\frac{2}{3}\mu_\text{eff}}
-6e^{-\frac{4}{3}\mu_\text{eff}}
+\biggl[-\frac{2\mu_\text{eff}^2+2\mu_\text{eff}+1}{3\pi^2}
+\frac{92}{9}\biggr]e^{-2\mu_\text{eff}}
-30e^{-\frac{8}{3}\mu_\text{eff}}
\nonumber\\&\quad
+\frac{1088}{15}e^{-\frac{10}{3}\mu_\text{eff}}
+\biggl[-\frac{3(8\mu_\text{eff}^2+4\mu_\text{eff}+1)}{8\pi^2}
-210\biggr]e^{-4\mu_\text{eff}}
+\frac{12160}{21}e^{-\frac{14}{3}\mu_\text{eff}}
\nonumber\\&\quad
+{\mathcal O}(e^{-\frac{16}{3}\mu_\text{eff}}),
\nonumber\\
J^\text{np}_{k=4}&=2\sqrt{2}e^{-\frac{1}{2}\mu_\text{eff}}
+\biggl[\frac{\mu_\text{eff}+1}{\pi}-4\biggr]e^{-\mu_\text{eff}}
+\frac{16\sqrt{2}}{3}e^{-\frac{3}{2}\mu_\text{eff}}
\nonumber\\&\quad
+\biggl[-\frac{7(2\mu_\text{eff}^2+2\mu_\text{eff}+1)}{4\pi^2}
-\frac{165}{8}\biggr]e^{-2\mu_\text{eff}}
+\frac{258\sqrt{2}}{5}e^{-\frac{5}{2}\mu_\text{eff}}
\nonumber\\&\quad
+\biggl[\frac{26(3\mu_\text{eff}+1)}{9\pi}-\frac{568}{3}\biggr]
e^{-3\mu_\text{eff}}
+\frac{2480\sqrt{2}}{7}e^{-\frac{7}{2}\mu_\text{eff}}
+{\mathcal O}(e^{-4\mu_\text{eff}}),
\nonumber\\
J^\text{np}_{k=6}&=\frac{8}{\sqrt{3}}e^{-\frac{1}{3}\mu_\text{eff}}
-\frac{14}{3}e^{-\frac{2}{3}\mu_\text{eff}}
+\biggl[\frac{2(\mu_\text{eff}+1)}{3\pi}+8\sqrt{3}\biggr]
e^{-\mu_\text{eff}}
-\frac{154}{3}e^{-\frac{4}{3}\mu_\text{eff}}
\nonumber\\&\quad
+\frac{1472}{5\sqrt{3}}e^{-\frac{5}{3}\mu_\text{eff}}
+\biggl[-\frac{7(2\mu_\text{eff}^2+2\mu_\text{eff}+1)}{6\pi^2}
-\frac{19427}{36}\biggr]e^{-2\mu_\text{eff}}
+\frac{6960\sqrt{3}}{7}e^{-\frac{7}{3}\mu_\text{eff}}
\nonumber\\&\quad
+{\mathcal O}(e^{-\frac{8}{3}\mu_\text{eff}}).
\end{align*}
\end{minipage}
}
\end{center}
\caption{Instanton expansion in the $(2,1)_k$ model in table \ref{21instanton} rewritten in terms of the effective chemical potential $\mu_\text{eff}$.}
\label{21mueff}
\end{table}

\subsubsection{Worldsheet instanton}\label{21ws}

Now let us proceed to the worldsheet instanton.
From the information on the position where we expect poles, it is not difficult to find
\begin{align}
d_1
=\frac{4\cos\frac{\pi}{k}}{\sin^2\frac{2\pi}{k}},\quad
d_2
=\frac{2\cos\frac{2\pi}{k}}{\sin^2\frac{4\pi}{k}}
-\frac{4+\cos\frac{2\pi}{k}}{\sin^2\frac{2\pi}{k}},\quad
d_3
=\frac{4\cos\frac{3\pi}{k}}{3\sin^2\frac{6\pi}{k}}
+\frac{12\cos\frac{\pi}{k}}{\sin^2\frac{2\pi}{k}}.
\label{d123}
\end{align}
Again it is easy to find the previously itemized properties in section \ref{22ws} still hold.
With the help of the multi-covering structure, we are able to write down higher instantons
\begin{align}
d_1&=\delta_1(k),&
\delta_1(k)&=\frac{4\cos\frac{\pi}{k}}{\sin^2\frac{2\pi}{k}},
\nonumber\\
d_2&=\frac{1}{2}\delta_1\Bigl(\frac{k}{2}\Bigr)
+\delta_2(k),&
\delta_2(k)&=-\frac{4+\cos\frac{2\pi}{k}}{\sin^2\frac{2\pi}{k}},
\nonumber\\
d_3&=\frac{1}{3}\delta_1\Bigl(\frac{k}{3}\Bigr)
+\delta_3(k),&
\delta_3(k)&=\frac{12\cos\frac{\pi}{k}}{\sin^2\frac{2\pi}{k}},
\nonumber\\
d_4&=\frac{1}{4}\delta_1\Bigl(\frac{k}{4}\Bigr)
+\frac{1}{2}\delta_2\Bigl(\frac{k}{2}\Bigr)
+\delta_4(k),&
\delta_4(k)
&=-\frac{32+16\cos\frac{2\pi}{k}}{\sin^2\frac{2\pi}{k}}+5,
\nonumber\\
d_5&=\frac{1}{5}\delta_1\Bigl(\frac{k}{5}\Bigr)
+\delta_5(k),&
\delta_5(k)&=
\frac{220\cos\frac{\pi}{k}+20\cos\frac{3\pi}{k}}{\sin^2\frac{2\pi}{k}}
-96\cos\frac{\pi}{k}.
\label{d45}
\end{align}
Besides, it is surprising to find the following property.
\begin{itemize}
\item
When the instanton number is small, if we replace the cosine functions in the numerators simply by $1$ and halve the argument of the sine functions in the denominators, the coefficient of the worldsheet instanton \eqref{d45} reduces to the expression of the $(2,2)_k$ model \eqref{d22123}, \eqref{d2245}.
\end{itemize}

\subsection{Higher worldsheet instantons}
In the previous two subsections we have studied the $(2,2)_k$ model and the $(2,1)_k$ model respectively.
In section \ref{22tst} we have noticed that the diagonal Gopakumar-Vafa invariants match with those of the local $D_5$ del Pezzo geometry.
Since the BPS indices themselves look different, one may suspect that the match is a mere coincidence.
Also, in section \ref{21ws} we have observed an interesting relation of the worldsheet instantons between the $(2,2)_k$ model and the $(2,1)_k$ model.
To study these observations in more details, we shall proceed to higher worldsheet instantons in this subsection.

\subsubsection{$(2,2)_k$ model}\label{2267}
In table \ref{mueff} we have studied the instanton expansion in the $(2,2)_k$ model up to ${\cal O}(e^{-\frac{16}{k}\mu_\text{eff}})$.
To fully utilize the data, we first note that, although we do not have the exact membrane instanton coefficients $\widetilde b_\ell$ for $\ell=6,7$, if we assume the multi-covering structure
\begin{align}
\widetilde b_6=\frac{1}{6}\beta_1(6k)+\frac{1}{3}\beta_2(3k)+\frac{1}{2}\beta_3(2k)+\beta_6(k),\quad
\widetilde b_7=\frac{1}{7}\beta_1(7k)+\beta_7(k),
\end{align}
and the trigonometric expression
\begin{align}
\beta_6(k)=\frac{\sum_{n=1}^{n_\text{max}}m_{6,n}\sin 2\pi kn}
{\pi\sin^2\pi k},\quad
\beta_7(k)=\frac{\sum_{n=1}^{n_\text{max}}m_{7,n}\sin 2\pi kn}
{\pi\sin^2\pi k},
\end{align}
the finite coefficients after the cancellation of divergences in
\begin{align}
\lim_{k\to 1}\bigl(d_m e^{-m\mu_\text{eff}/k}
+(\widetilde b_\ell\mu_\text{eff}
+\widetilde c_\ell)
e^{-\ell\mu_\text{eff}}\bigr),
\end{align}
with \eqref{22der} only depend on the linear combination of $m_{\ell,n}$ which is determined from the first two terms in the WKB expansion.
This is due to the periodicity of the trigonometric functions.
Using the terms of $e^{-\frac{12}{k}\mu_\text{eff}}$ in $J^\text{np}_k$ ($k=1,2,3,4,6$), we can find the sixth worldsheet instanton coefficient.
Similarly, the seventh worldsheet instanton coefficient is found from the terms of $e^{-\frac{14}{k}\mu_\text{eff}}$.
The results are
\begin{align}
d_6&=\frac{1}{6}\delta_1\Bigl(\frac{k}{6}\Bigr)
+\frac{1}{3}\delta_2\Bigl(\frac{k}{3}\Bigr)
+\frac{1}{2}\delta_3\Bigl(\frac{k}{2}\Bigr)
+\delta_6(k),\nonumber\\
&\qquad\delta_6(k)
=-\frac{1359}{\sin^2\frac{\pi}{k}}+1280
-320\sin^2\frac{\pi}{k},\nonumber\\
d_7&=\frac{1}{7}\delta_1\Bigl(\frac{k}{7}\Bigr)
+\delta_7(k),\nonumber\\
&\qquad\delta_7(k)
=\frac{8428}{\sin^2\frac{\pi}{k}}-14816
+10048\sin^2\frac{\pi}{k}-2560\sin^4\frac{\pi}{k}.
\label{22ws67}
\end{align}
These can be summarized into the diagonal Gopakumar-Vafa invariants.
See table \ref{GV} of $d=6,7$.
We observe that the match of the diagonal Gopakumar-Vafa invariants between the $(2,2)_k$ model and the local $D_5$ del Pezzo geometry \cite{KKV} still holds for higher instantons.

\subsubsection{More numerical data for $(2,1)_k$ model}

To study higher worldsheet instantons in the $(2,1)_k$ model, we need more numerical data.
Besides $k=1,2,3,4,6$ which we have studied in table \ref{21mueff}, the simplest cases are probably $k=5$, $k=8$ and $k=12$.
The study of these cases is important also because the coefficients of the worldsheet instantons we have found in section \ref{21ws} are rather complicated and this extra information provides non-trivial checks to them. 

We have found the exact values of the partition function $Z_k(N)$ up to $N_\text{max}$ for $(k,N_\text{max})=(5,4),(8,5),(12,7)$.
The first few values for $k=8$ and $k=12$ are listed in table \ref{morevalues}, while those for $k=5$ were listed in \cite{MN1}.
\begin{table}[ht]
\begin{center}
\framebox{
\begin{minipage}{0.9\textwidth}
\begin{align*}
&Z_{8}(1)=\frac{1}{32\pi},\quad
Z_{8}(2)=\frac{-224+(85-44\sqrt{2})\pi^2}{65536\pi^2},\nonumber\\
&\quad Z_{8}(3)=\frac{-11040+(9649-60\sqrt{2})\pi^2
-90(3+19\sqrt{2})\pi^3}{94371840\pi^3},
\nonumber\\
&Z_{12}(1)=\frac{1}{48\pi},\quad
Z_{12}(2)=\frac{-14256+(12919-6624\sqrt{3})\pi^2}{5971968\pi^2},
\nonumber\\
&\quad Z_{12}(3)=\frac{-2041200+(3488481-272160\sqrt{3})\pi^2
-20(38727+3464\sqrt{3})\pi^3}{38698352640\pi^3}.
\end{align*}
\end{minipage}
}
\end{center}
\caption{More exact values of the partition function $Z_k(N)$ for the $(2,1)_k$ model.}
\label{morevalues}
\end{table}
Then, we can compare our expectations from \eqref{d45}, \eqref{cb} with these exact values and proceed to find higher instanton coefficients by fitting the values.
The results in terms of $\mu_\text{eff}$ are summarized in table \ref{58Ceff}.
\begin{table}[ht]
\begin{center}
\framebox{
\begin{minipage}{0.9\textwidth}
\begin{align*}
J^\text{np}_{k=5}&=\frac{8}{\sqrt{5}}e^{-\frac{2}{5}\mu_\text{eff}}
+\biggl[\frac{9}{\sqrt{5}}-7\biggr]e^{-\frac{4}{5}\mu_\text{eff}}
+\frac{64}{3\sqrt{5}}e^{-\frac{6}{5}\mu_\text{eff}}
+\biggl[\frac{23}{2\sqrt{5}}-\frac{93}{2}\biggr]
e^{-\frac{8}{5}\mu_\text{eff}}
\nonumber\\&\quad
+\biggl[-\frac{2\mu_\text{eff}^2+2\mu_\text{eff}+1}{5\pi^2}+52\sqrt{5}
\biggr]e^{-2\mu_\text{eff}}
+\biggl[\frac{232}{\sqrt{5}}-\frac{1246}{3}\biggr]
e^{-\frac{12}{5}\mu_\text{eff}}
\nonumber\\&\quad
+\biggl[\frac{18584}{7\sqrt{5}}-312\biggr]
e^{-\frac{14}{5}\mu_\text{eff}}
+{\mathcal O}(e^{-\frac{16}{5}\mu_\text{eff}}),
\nonumber\\
J^\text{np}_{k=8}&=4\sqrt{2+\sqrt{2}}e^{-\frac{1}{4}\mu_\text{eff}}
-8e^{-\frac{1}{2}\mu_\text{eff}}
+\frac{4}{3}(8+\sqrt{2})\sqrt{2+\sqrt{2}}
e^{-\frac{3}{4}\mu_\text{eff}}
\nonumber\\&\quad
+\biggl[\frac{\mu+1}{2\pi}-(61+16\sqrt{2})\biggr]
e^{-\mu_\text{eff}}
+\frac{4}{5}(191+24\sqrt{2})\sqrt{2+\sqrt{2}}
e^{-\frac{5}{4}\mu_\text{eff}}
\nonumber\\&\quad
-\frac{32}{3}(79+33\sqrt{2})e^{-\frac{3}{2}\mu_\text{eff}}
+\frac{4}{7}(3576+973\sqrt{2})\sqrt{2+\sqrt{2}}
e^{-\frac{7}{4}\mu_\text{eff}}
+{\mathcal O}(e^{-2\mu_\text{eff}}),
\nonumber\\
J^\text{np}_{k=12}&=4\sqrt{2}(1+\sqrt{3})e^{-\frac{1}{6}\mu_\text{eff}}
-\frac{2}{3}(24+\sqrt{3})e^{-\frac{1}{3}\mu_\text{eff}}
+\frac{2}{3}\sqrt{2}(19+18\sqrt{3})e^{-\frac{1}{2}\mu_\text{eff}}
\nonumber\\&\quad
-\frac{8}{3}(47+12\sqrt{3})e^{-\frac{2}{3}\mu_\text{eff}}
+\frac{24}{5}\sqrt{2}(49+41\sqrt{3})e^{-\frac{5}{6}\mu_\text{eff}}
\nonumber\\&\quad
+\biggl[\frac{\mu_\text{eff}+1}{3\pi}
-\frac{2}{3}(3506+1383\sqrt{3})\biggr]e^{-\mu_\text{eff}}
+\frac{8}{7}\sqrt{2}(5387+3860\sqrt{3})e^{-\frac{7}{6}\mu_\text{eff}}
\nonumber\\&\quad
+{\mathcal O}(e^{-\frac{4}{3}\mu_\text{eff}}),
\end{align*}
\end{minipage}
}
\end{center}
\caption{Instanton expansion in the $(2,1)_k$ model for $k=5,8,12$ in terms of the effective chemical potential $\mu_\text{eff}$.}
\label{58Ceff}
\end{table}
See table \ref{fit} for the comparison of these coefficients with the numbers found by fitting the exact values.
\begin{table}
\begin{center}
\begin{tabular}{|c|c|r|r|}
\hline
&&numerical values& expected exact values\\
\hline\hline
$k=5$&$e^{-\frac{2}{5}\mu_\text{eff}}$
&$3.57770877400$&$8/\sqrt{5}\simeq\phantom{-}3.57770876400$\\
&$e^{-\frac{4}{5}\mu_\text{eff}}$
&$-2.97507762647$&$9/\sqrt{5}-7\simeq-2.97507764050$\\
&$e^{-\frac{6}{5}\mu_\text{eff}}$
&$9.54055670898$&$64/(3\sqrt{5})\simeq\phantom{-}9.54055670400$\\
&$e^{-\frac{8}{5}\mu_\text{eff}}$
&$-41.3570436490$&$23/(2\sqrt{5})-93/2\simeq-41.3570436518$\\
&$e^{-2\mu_\text{eff}}$
&$116.275534828$&$52\sqrt{5}\simeq\phantom{-}116.275534830$\\
&$e^{-\frac{12}{5}\mu_\text{eff}}$
&$-311.579779177$&$232/\sqrt{5}-1246/3\simeq-311.579779177$\\
&$e^{-\frac{14}{5}\mu_\text{eff}}$
&$875.288208396$&$18584/(7\sqrt{5})-312\simeq\phantom{-}875.288208396$\\
\hline\hline
$k=8$&$e^{-\frac{1}{4}\mu_\text{eff}}$
&$7.39103628224$&$4\sqrt{2+\sqrt{2}}\simeq\phantom{-}7.39103626009$\\
&$e^{-\frac{1}{2}\mu_\text{eff}}$
&$-7.99999998763$&$-8\simeq-8.00000000000$\\
&$e^{-\frac{3}{4}\mu_\text{eff}}$
&$23.1935979756$&$4\sqrt{2+\sqrt{2}}(8+\sqrt{2})/3
\simeq\phantom{-}23.1935979332$\\
&$e^{-\mu_\text{eff}}$
&$-83.6274170573$&$-(61+16\sqrt{2})\simeq-83.6274169980$\\
&$e^{-\frac{5}{4}\mu_\text{eff}}$
&$332.509602923$&$4\sqrt{2+\sqrt{2}}(191+24\sqrt{2})/5\simeq\phantom{-}332.509602987$\\
&$e^{-\frac{3}{2}\mu_\text{eff}}$
&$-1340.46984066$&$-32(79+33\sqrt{2})/3\simeq-1340.46984062$\\
&$e^{-\frac{7}{4}\mu_\text{eff}}$
&$5228.66168352$&$4\sqrt{2+\sqrt{2}}(3576+973\sqrt{2})/7
\simeq\phantom{-}5228.66168353$\\
\hline\hline
$k=12$&$e^{-\frac{1}{6}\mu_\text{eff}}$
&$15.4548133432$&$4\sqrt{2}(1+\sqrt{3})\simeq\phantom{-}15.4548132206$\\
&$e^{-\frac{1}{3}\mu_\text{eff}}$
&$-17.1547004735$&$-2(24+\sqrt{3})/3\simeq-17.1547005384$\\
&$e^{-\frac{1}{2}\mu_\text{eff}}$
&$47.3072487510$&$2\sqrt{2}(19+18\sqrt{3})/3\simeq\phantom{-}47.3072487035$\\
&$e^{-\frac{2}{3}\mu_\text{eff}}$
&$-180.758959157$&$-8(47+12\sqrt{3})/3\simeq-180.758959176$\\
&$e^{-\frac{5}{6}\mu_\text{eff}}$
&$814.682611241$&$24\sqrt{2}(49+41\sqrt{3})/5\simeq\phantom{-}814.682611250$\\
&$e^{-\mu_\text{eff}}$
&$-3934.28417792$&$-2(3506+1383\sqrt{3})/3\simeq-3934.28417791$\\
&$e^{-\frac{7}{6}\mu_\text{eff}}$
&$19512.4558488$&$8\sqrt{2}(5387+3860\sqrt{3})/7\simeq\phantom{-}19512.4558487$\\
\hline
\end{tabular}
\end{center}
\caption{Comparison of numerical values obtained from fitting and expected exact values for the non-perturbative coefficients of $e^{-\frac{2m}{k}\mu_\text{eff}}$ in table \ref{58Ceff}.}
\label{fit}
\end{table}

\subsubsection{$(2,1)_k$ model}\label{higher21}
Having obtained some extra exact values, let us now proceed to obtain the function form of higher worldsheet instanton coefficients in the $(2,1)_k$ model.
Note that, although we do not know the sixth and seventh membrane instanton coefficients we can use the data of $k=2$ as well due to the reason explained in section \ref{2267}.
Also, if we assume the coefficients of cosine functions to be rational numbers, the conditions from the $k=5$ case, the $k=8$ case and the $k=12$ case give two relations respectively.
Hence we can fully determine the coefficients.
Using the data at $k=2,3,4,6$ in table \ref{21mueff} and $k=5,8,12$ in table \ref{58Ceff}, we find
\begin{align}
d_6&=\frac{1}{6}\delta_1\Bigl(\frac{k}{6}\Bigr)
+\frac{1}{3}\delta_2\Bigl(\frac{k}{3}\Bigr)
+\frac{1}{2}\delta_3\Bigl(\frac{k}{2}\Bigr)
+\delta_6(k),\nonumber\\
&\qquad
\delta_6(k)=-\frac{780+579\cos\frac{2\pi}{k}}{\sin^2\frac{2\pi}{k}}
+\Bigl(848+480\cos\frac{2\pi}{k}\Bigr)
-\Bigl(256+64\cos\frac{2\pi}{k}\Bigr)\sin^2\frac{2\pi}{k},\nonumber\\
d_7&=\frac{1}{7}\delta_1\Bigl(\frac{k}{7}\Bigr)
+\delta_7(k),\nonumber\\
&\qquad\delta_7(k)
=\frac{7168\cos\frac{\pi}{k}+1260\cos\frac{3\pi}{k}}
{\sin^2\frac{2\pi}{k}}
-\Bigl(13232\cos\frac{\pi}{k}+1696\cos\frac{3\pi}{k}\Bigr)
\nonumber\\
&\qquad\qquad\qquad
+\Bigl(9472\cos\frac{\pi}{k}+576\cos\frac{3\pi}{k}\Bigr)
\sin^2\frac{2\pi}{k}-2560\cos\frac{\pi}{k}\sin^4\frac{2\pi}{k}.
\label{21ws67}
\end{align}

Now let us compare the worldsheet instanton coefficients \eqref{21ws67} in the $(2,1)_k$ model with the worldsheet instanton coefficients \eqref{22ws67} in the $(2,2)_k$ model.
If we apply the rule we have found in section \ref{21ws} to \eqref{21ws67}, we find
\begin{align}
\delta^{(2,1)\to(2,2)}_6(k)&=-\frac{1359}{\sin^2\frac{\pi}{k}}
+1328-320\sin^2\frac{\pi}{k},\nonumber\\
\delta^{(2,1)\to(2,2)}_7(k)&=\frac{8428}{\sin^2\frac{\pi}{k}}
-14928+10048\sin^2\frac{\pi}{k}-2560\sin^4\frac{\pi}{k},
\end{align}
which is very close to \eqref{22ws67} but contains some discrepancies.
Our analysis here can be summarized as follows.
\begin{itemize}
\item
The relation of the worldsheet instantons between the $(2,2)_k$ model and the $(2,1)_k$ model observed in section \ref{21ws} is mostly valid for higher instanton numbers, though a modification should be taken into account.
\end{itemize}

\subsubsection{Towards topological string theory}\label{21top}
Finally, let us make some efforts to guess an expression for the $(2,1)_k$ model, that is similar to the free energy of the topological string theory \eqref{Jtop}.

Although in the study of higher worldsheet instantons in section \ref{higher21} we have found some discrepancies, since the relation mostly holds, let us neglect the discrepancies shortly and try to derive possible conclusions out of the relation between the $(2,2)_k$ model and the $(2,1)_k$ model observed in section \ref{21ws}.
When we say that after some procedures of the replacements the coefficients of the worldsheet instantons in the $(2,1)_k$ model reduce to those in the $(2,2)_k$ model, we come up with three possibilities.\footnote{We are grateful to Kazumi Okuyama for valuable discussions.}
\begin{itemize}
\item[$\circ$]
One possibility is the imaginary part of the K\"{a}hler parameter.
When we relate the chemical potential of the $(2,2)$ model to the K\"{a}hler parameter we do not need to introduce the imaginary part.
However, for the $(2,1)$ model, the cosine functions can come in because of the imaginary part.
\item[$\circ$]
Another possibility is that the $(2,1)_k$ model shares the same BPS index with the $(2,2)_k$ model, but is different in the function forms of the free energy.
Due to the information of the BPS index it picks up different linear combinations of the cosine functions.
\item[$\circ$]
The last possibility is that the topological invariants of the $(2,1)_k$ model have more refined structures to distinguish two types of arguments in the cosine functions than those of the $(2,2)_k$ model.
\end{itemize}
In any case, we do not have a clear geometric picture and we cannot make a concrete decision out of these possibilities.
For the first possibility, if there are many enough K\"{a}hler parameters, we may assign different imaginary parts to reproduce the cosine functions.
Still since in the ABJM case the imaginary part comes in by shifting the K\"ahler parameters by $\pm\pi i$ instead of the chemical potentials, it seems difficult to obtain the $k^{-1}$ dependence in the arguments of the cosine functions.
For the last possibility, though it is interesting to lift the topological invariants of the $(2,2)_k$ model to more refined structures in the $(2,1)_k$ model, since we do not know either the BPS index or the free energy function, we have little to say on this possibility.
We have concentrated on the second possibility and found an expression consistent with the relation of the replacements, the identification of the BPS indices and the requirement of the pole cancellation mechanism.
However, due to the lack of data we are not sure of this proposal.

\section{Discussion}\label{discuss}

In this paper we have found that the partition function of superconformal Chern-Simons theories, other than the ABJM matrix model, can also be described by the free energy of the refined topological string theory or its deformation. 
For the $(2,2)_k$ model we find that the instanton expansion matches well with the ABJM case.
Namely we can use the same function obtained from the free energy of the refined topological string theory to describe the grand potential of the $(2,2)_k$ model.
The only differences appear in the set of topological invariants and in the identification of the K\"{a}hler parameter $T$ and the string coupling $g_s$ with the chemical potential $\mu$ and the level $k$.
For the $(2,1)_k$ model the situation is more obscure.
After observing the similarity of the worldsheet instantons between the $(2,2)_k$ model and the $(2,1)_k$ model itemized in section \ref{21ws} and section \ref{higher21}, we have proposed several possibilities for the  instanton coefficients. 
However, we cannot determine the full instanton expansion.

Let us discuss several future directions.
Apparently, it is a crucial question to identify the Calabi-Yau manifold which carries the topological invariants of the $(2,2)_k$ model.
It is interesting to observe that the diagonal Gopakumar-Vafa invariants of the $(2,2)_k$ model match with those of the local $D_5$ del Pezzo geometry \cite{KKV} at least up to the seventh instanton, though the BPS indices look different \cite{HKP}.
If this match holds for all instanton numbers, it may suggest that there are two different manifolds sharing the same diagonal Gopakumar-Vafa invariants, which is surprising to us.  
Since our analysis only gives the diagonal topological invariants at a fixed number of $d=\sum_i{\bf d}_i$, it is desirable to see the deformation with different ranks.
Here we expect that either the formulation of \cite{AHS,H,HO} or the formulation of \cite{MM} for the ABJ theory \cite{HLLLP1,ABJ} with the gauge group $U(N_1)_k\times U(N_2)_{-k}$ would be applicable to this deformation.
Also, in \cite{CM} the standard computation of the WKB expansion was simplified to the semiclassical TBA techniques.
It is interesting to see how these techniques work for general $(q,p)_k$ models including the $(2,2)_k$ model.
After generating more terms in the $k$ expansion with these techniques, we are expecting to find more topological invariants.
It is also interesting to observe that the perturbative coefficient $C$ and the exponential factor of the membrane instanton $e^{-\mu_\text{eff}}$ coincide with those of the quantum mechanical model associated to local ${\cal B}_3$ \cite{GHM1}. See also \cite{KM,HW,WWH,GHM2}.

The relation of the worldsheet instantons between the $(2,2)_k$ model and the $(2,1)_k$ model is also very interesting.
This relation may be helpful to determine the instanton expansion of the $(2,1)_k$ model.
It would be an interesting future direction to proceed to higher instanton numbers to gain more information to study the relation and determine the full instanton expansion.

As the $(2,2)_k$ and $(2,1)_k$ models are very special cases, it is interesting to see whether there is an unexpected symmetry enhancement in these models.

Though we have restricted ourselves to the $(2,2)_k$ model and the $(2,1)_k$ model, it would also be interesting to extend our study to the general $(q,p)_k$ model.
After that, we hope to go beyond the ``minimal'' $(q,p)_k$ cases to study those without the constraint \eqref{sofminimal}.
The result is expected to be more complicated, due to the possible non-trivial dependence on the ordering.
However, the result is explicitly known \cite{HM} for the case where the quiver is given as a repetition of that of the ABJM theory \cite{GW,HLLLP1}, which will provide some hints in this direction.

\section*{Acknowledgements}
We are grateful to Heng-Yu Chen, Yasuyuki Hatsuda, Hiroaki Kanno, Marcos Mari\~no and Kazumi Okuyama for valuable discussions.
We are also grateful to the participants of the 7th Taiwan String Workshop where the results of this paper were presented by S.M.
The work of S.M.\ is supported by JSPS Grant-in-Aid for Scientific
Research (C) \# 26400245, while the work of T.N.\ is partly supported
by the JSPS Research Fellowships for Young Scientists.

\end{document}